# *Size effects and idealized dislocation microstructure at small scales: predictions of a phenomenological model of Mesoscopic Field Dislocation Mechanics: Part II*

Anish Roy and Amit Acharya[*]

Civil and Environmental Engineering, Carnegie Mellon University, Pittsburgh, PA 15213, U.S.A

**Abstract**

In Part I of this set of two papers, a model of mesoscopic plasticity is developed for studying initial-boundary value problems of small scale plasticity. Here we make qualitative, finite element method-based computational predictions of the theory. We demonstrate size effects and the development of strong inhomogeneity in simple shearing of plastically-constrained grains. Nonlocality in elastic straining leading to a strong Bauschinger effect is analyzed. Low shear strain boundary layers in constrained simple shearing of infinite layers of polycrystalline materials are not predicted by the model, and we justify the result based on an examination of the no-dislocation-flow boundary condition in the model. The time dependent, spatially homogeneous, simple shearing solution of PMFDM is studied numerically. The computational results and an analysis of continuous dependence with respect to initial data of solutions for a model linear problem point to the need for a nonlinear study of a stability transition of the homogeneous solution with decreasing grain size and increasing applied deformation. The continuous-dependence analysis also points to a possible mechanism for the development of spatial inhomogeneity in the initial stages of deformation in lower-order gradient plasticity theory. Results from thermal cycling of small scale beams/films with different degrees of constraint to plastic flow are presented showing size effects and reciprocal-film-thickness scaling of dislocation density boundary layer width. Qualitative similarities with results from discrete dislocation analyses are noted where possible.

We discuss the convergence of approximate solutions with mesh refinement and its implications for the prediction of dislocation microstructure development, motivated by the notion of measure-valued solutions to conservation laws.

## 1. Introduction

This work presents results of a finite-element approximation of Phenomenological Mesoscopic Field Dislocation Mechanics (PMFDM), introduced in Part I of this paper (Acharya and Roy, 2005). Here, mesoscopic plasticity is modeled as an extension of conventional plasticity, while accounting for the effects of dislocation stresses as well as their spatio-temporal evolution in a physically meaningful averaged sense. An important characteristic of PMFDM is that it requires only one addition material parameter over conventional plasticity.

The paper is organized as follows: in Section 2, a brief, self-contained description of the governing relations of PMFDM is presented for the reader not interested in all details of the development. Section 3 comprises the finite element discretization strategy for the model. In

---

[*] Corresponding author: Tel. (412) 268 4566; Fax. (412) 268 7813; email: acharyaamit@cmu.edu

Section 4, results of the implementation are presented and discussed. The paper ends with some concluding remarks in Section 5.

## 2. PMFDM

For the convenience of the reader, the formulation of PMFDM (Acharya and Roy, 2005) is briefly summarized below. The equations prescribing the elastic incompatibility are

$$\begin{aligned} curl\,\chi &= \alpha \\ div\,\chi &= \mathbf{0} \\ \chi \mathbf{n} &= \mathbf{0} \text{ on } \partial B. \end{aligned} \quad (1)$$

Here, $\chi$ is the incompatible part of the elastic distortion tensor $U^e$, $\mathbf{n}$ is the unit normal on the boundary of the body $\partial B$ and $\alpha$ is the dislocation density tensor. The compatible part of $U^e$ is given by $grad\,(\mathbf{u}-\mathbf{z})$, where $\mathbf{u}$ is the total displacement field and $\mathbf{z}$ obeys the relation,

$$div\big(grad\,\dot{\mathbf{z}}\big) = div\big(\alpha \times V + L^p\big). \quad (2)$$

$V$, the averaged dislocation velocity tensor, and $L^p$ need to be specified constitutively. Roughly speaking, $L^p$ has the physical meaning of representing that part of the total plastic strain rate not represented by the slipping produced by the averaged dislocation density. The value of $\dot{\mathbf{z}}$ is prescribed at an arbitrarily chosen point of the body and in our case assumed to vanish without loss of generality. The (symmetric) stress tensor $T$ satisfies

$$\begin{aligned} T &= C : (grad(\mathbf{u}-\mathbf{z}) + \chi) \\ div\,T &= \mathbf{0} \end{aligned} \quad (3)$$

where $C$ is the possibly anisotropic fourth order tensor of linear elastic moduli. Standard traction/displacement boundary conditions are to be used with the above equation. Finally the temporal evolution of the dislocation density tensor field is prescribed as

$$\dot{\alpha} = -curl\,(S) + s, \quad (4)$$

where $s$ is the dislocation nucleation rate tensor to be specified constitutively and $S$, the macroscopic slipping distortion, is defined as

$$S := \alpha \times V + L^p. \quad (5)$$

The least restrictive boundary condition on (4) would be to impose $\alpha(V \cdot \mathbf{n})$ on inflow points of the boundary (where $V \cdot \mathbf{n} < 0$), with a specification of $L^p \times \mathbf{n}$ on the entire boundary.

Initial conditions on the averaged field variables $\mathbf{u}$, $\alpha$, and $grad\,z$ are required. Given an initial specification of the $\alpha\big|_{t=0}$ field, the internal stress field at the initial instant is calculated and also the initial condition on *grad z* by solving (1), (3) with $\mathbf{u}|_{t=0} \equiv \mathbf{0}$ and $z$ specified at one single point arbitrarily.

The constitutive choices of the elastic moduli $\mathbf{C}$ and the slipping distortion (5) introduce quantities that we model phenomenologically to complete the model. Simple choices motivated by conventional plasticity and the thermodynamics of PMFDM are

$$\mathbf{L}^p = \dot{\gamma}\frac{\mathbf{T}'}{|\mathbf{T}'|} \quad ; \quad \dot{\gamma} \geq 0,$$

$$\mathbf{V} = v\frac{\mathbf{d}}{|\mathbf{d}|} \quad ; \quad v \geq 0, \tag{6}$$

where, $\mathbf{T}'$ is the stress deviator, $\dot{\gamma}$ and $v$ are non-negative functions of state representing the magnitudes of the SSD slipping rate and the averaged dislocation velocity respectively. The direction of the dislocation velocity is defined by

$$\mathbf{d} := \mathbf{b} - \left(\mathbf{b}\cdot\frac{\mathbf{a}}{|\mathbf{a}|}\right)\frac{\mathbf{a}}{|\mathbf{a}|},$$

$$\mathbf{b} := X(\mathbf{T}'\boldsymbol{\alpha}) \quad ; \quad b_i = e_{ijk}T'_{jr}\alpha_{rk} \quad ; \quad \mathbf{a} := X(tr(\mathbf{T})\boldsymbol{\alpha}) \quad ; \quad a_i = \left(\frac{1}{3}T_{mm}\right)e_{ijk}\alpha_{jk}. \tag{7}$$

We choose a power law relation for $\dot{\gamma}$ as

$$\dot{\gamma} = \dot{\gamma}_0\left(\frac{|\mathbf{T}'|}{\sqrt{2}g}\right)^{\frac{1}{m}}, \tag{8}$$

where $m$ is the macroscopic rate-sensitivity of the material, $g$ is the strength of the material, and $\dot{\gamma}_0$ is a reference strain rate. The expression for $v$ is assumed to be

$$v(state) = \eta^2 b\left(\frac{\mu}{g}\right)^2 \dot{\gamma}(\mathbf{T}', g), \tag{9}$$

where $\mu$ is the shear modulus, $b$ the Burgers vector magnitude and $\eta = 1/3$ a material parameter. For further motivation behind these choices, refer to Part I of this paper. For the purpose of the current study of PMFDM we do not consider the effect of dislocation nucleation, i.e. $\mathbf{s} \equiv \mathbf{0}$.

The strength of the material is assumed to evolve according to

$$\dot{g} = \left[\frac{\eta^2\mu^2 b}{2(g-g_0)}k_0|\boldsymbol{\alpha}| + \theta_0\left(\frac{g_s - g}{g_s - g_0}\right)\right]\{|\boldsymbol{\alpha}\times\mathbf{V}| + \dot{\gamma}\}, \tag{10}$$

where $g_s$ is the saturation stress, $g_0$ is the yield stress, and $\theta_0$ is the Stage II hardening rate. It is to be noted that the material parameters $g_s, g_0, \mu, b, \dot{\gamma}_0, m$ are part of the conventional Voce Law prescription for strength. The only extra parameter that needs to be experimentally fitted to the material under consideration is $k_0$. The addition of Field Dislocation Mechanics (FDM), essential for the accurate prediction of internal stress fields, adds no extra parameters to the PMFDM formulation.

*Initial Conditions:*

The field equations mentioned above admits initial conditions on the fields $\boldsymbol{u}$, $\boldsymbol{\alpha}$ and $grad\,\boldsymbol{z}$ which are as follows. For the $\boldsymbol{u}$ field we assume $\boldsymbol{u}|_{t=0} \equiv \boldsymbol{0}$, which is a physically natural initial condition on the displacement field. Unless otherwise mentioned, we assume that the body is initially dislocation free which translates to $\boldsymbol{\alpha}|_{t=0} \equiv \boldsymbol{0}$. Initial condition on the $grad\,\boldsymbol{z}$ field is obtained from solving (1), (3) with $\boldsymbol{u}|_{t=0} \equiv \boldsymbol{0}$ and the value of $\boldsymbol{z}$ set to zero at a single arbitrary point in the body.

## 3. Finite Element Discretization

The finite element discretization for the system of equations in Section 2 is similar to the FDM discretization in Roy and Acharya (2005). Here we only summarize those equations which have a slightly different formulation.

In the following, the symbol $\delta(\cdot)$ represents a variation (or test function) associated with the field $(\cdot)$ in a suitable class of functions. An increment of time $[t, t+\Delta t]$ is considered, and fields without any superscripts refer to values at $t + \Delta t$ and those with the superscript $t$ refer to values at time $t$. All spatial fields are discretized by first-order, 8-node (three-dimensional), isoparametric brick elements.

The discretization for (2) is

$$\int_B \delta z_{i,j} \left[ z_{i,j} - z_{i,j}^t - \Delta t \left( L_{ij}^{p\,t} + e_{jmk} \alpha_{im}^t V_k^t \right) \right] dv = 0 \;; \tag{11}$$

specify $z_i = 0$ at an arbitrarily chosen point.

As usual, (11) holds for all choices of $\delta z_i$ consistent with the above mentioned kinematic constraint.

A mixed Forward-Backward Euler scheme is adopted for (4) as

$$\int_B \delta\alpha_{ij}\left(\alpha_{ij} - \alpha_{ij}^t\right)dv - \Delta t \int_B \left[\delta\alpha_{ij,k}\alpha_{ij}v_k^t - \delta\alpha_{ij,k}\alpha_{ik}v_j^t\right]dv$$
$$- \Delta t \int_B \delta\alpha_{ij}s_{ij}^t \, dv + \Delta t \int_{\partial B_i} \delta\alpha_{ij}F_{ij} \, da$$
$$+ \Delta t \int_{\partial B_o} \delta\alpha_{ij}\alpha_{ij}^t\left(v_k^t n_k\right)da - \Delta t \int_{\partial B} \delta\alpha_{ij}\alpha_{ik}^t n_k v_j^t \, da \quad (12)$$
$$- \Delta t \int_B \underline{\delta\alpha_{ij,k} e_{jkl} L_{il}^p} \, dv + \Delta t \int_{\partial B} \underline{\delta\alpha_{ij} e_{jkl} L_{il}^p n_k} \, da$$
$$+ \int_{B_{interiors}} A_{ri}\left(\delta\alpha_{ri} + \Delta t\left[\delta\alpha_{ri,j}v_j^t + \delta\alpha_{ri}v_{j,j}^t - \delta\alpha_{rj,j}v_i^t - \delta\alpha_{rj}v_{i,j}^t\right]\right)dv = 0,$$

where

$$A_{ri} = \alpha_{ri} - \alpha_{ri}^t + \Delta t\left[\alpha_{ri,j}^t v_j^t + \alpha_{ri}^t v_{j,j}^t - \alpha_{rj,j}^t v_i^t - \alpha_{rj}^t v_{i,j}^t - s_{ri}^t + \underline{e_{ijk} L_{rk,j}^p}\right], \quad (13)$$

$F$ is the prescribed flux on the inflow boundary, $\partial B_o$ is the set of outflow/neutral points of the boundary where $V \cdot n \geq 0$ and $B_{interiors}$ refers to the union of the element interiors. Again, $\delta\alpha_{ij}$ is arbitrary up to satisfying any prescribed essential boundary conditions. The underlined terms in (12)-(13) are the additional terms that enter the discretization for the dislocation density evolution in PMFDM over FDM.

The time step is controlled by

$$\Delta t \leq \min\left\{f \frac{h}{|V|}, \frac{0.002}{|\boldsymbol{\alpha} \times V| + \dot{\gamma}}\right\}, \quad f \sim 0.1. \quad (14)$$

This reflects a conservative choice between a Courant condition and a maximum bound of 0.2% on the plastic strain increment. Here $h$ is a minimum element edge length.

## 4. Results and Discussion

The formulation in Section 3 is implemented in a Fortran code which invokes sequential PETSc (Balay et al., 2001) libraries. A problem is typically solved in steps. The first step solves the problem of internal stress due to the presence of a prescribed initial dislocation density in the body, and defines the initial condition for slip distortion. A time evolving analysis may be performed in the subsequent steps. The system of equations to be solved is broken up into parts. First $\bar{\alpha}$ is solved for with $S$ treated as data, followed by $\bar{\chi}$ and $\bar{z}$ where $\bar{\alpha}$ and $S$ are treated as data. Finally we solve for $\bar{u}$ with $\bar{\chi}$ and $grad\,\bar{z}$ treated as data.

Material parameters representative of Al is used for all the computational experiments except for $k_0$, for which we make the arbitrary choice of $k_0 = 20.0$ motivated by Acharya and Beaudoin (2000). The other material parameters used are $b = 4.05 \times 10^{-4}\,\mu m$, $m = 0.03$, $g_s = 161\,\text{MPa}$, $g_0 = 17.3\,\text{MPa}$ and $\theta_0 = 392.5\,\text{MPa}$. The coefficient of thermal expansion, $\vartheta$, is $23.5 \times 10^{-6} \text{K}^{-1}$. Isotropic elastic constants of the material are $E = 62.78\,\text{GPa}$, $\nu = 0.3647$, where $E$ is the Young's modulus and $\nu$ is the Poisson's ratio. The cubic crystal reference frame coincides with the rectangular Cartesian reference frame for the analysis. Fig. 1 is a schematic of a typical geometry that is used. Unless otherwise mentioned the reference strain rate is $\dot{\gamma}_0 = 1\,\text{sec}^{-1}$. In the interpretation of results, the symbol $\varepsilon$ represents the symmetric part of $\partial \boldsymbol{u}/\partial \boldsymbol{x}$.

All computations were performed on one of two desktop machines with 1GB and 2GB RAM, respectively.

## 4.1 Simple Shear of Constrained Grains

### 4.1.1 Size Effect

There is considerable experimental evidence that plastic flow in crystalline materials is size dependent over length scales of 100μm or less (Stölken and Evans, 1998; Ma and Clarke, 1995).

A regular mesh of $32 \times 32 \times 1$ elements is used to discretized four blocks of dimensions $(0.5\mu m)^3$, $(1\mu m)^3$, $(10\mu m)^3$ and $(100\mu m)^3$. These blocks are thought of as idealized grains. The imposed initial conditions are as in Section 2. The imposed boundary conditions are as follows: The grains are plastically constrained, which translates to a zero surface flow boundary condition. This is achieved by imposing

$$\boldsymbol{S} \times \boldsymbol{n} = \boldsymbol{0} \text{ on the surface.} \tag{15}$$

Thus all the external surfaces act as a rigid boundary to slipping.

The displacements on the bottom face are constrained in all three directions while those on the top, left and right faces are constrained in the $x_2, x_3$ directions only. The front and back faces are traction free. The displacements corresponding to a simple shear strain of 0.8% is prescribed through the kinematic boundary condition

$$u_1(x_1, x_2, x_3, t) = d(x_2) \dot{\Gamma} t \tag{16}$$

on the nodes of the left, right, top and bottom faces. Here, $d(x_2)$ is the height, from the bottom of the block, of the point with coordinates $(x_1, x_2, x_3)$ (Fig. 1). $\Gamma$ is the average engineering shear strain given by the ratio of the applied horizontal displacement of the top surface to the cube height, $\dot{\Gamma}$ is an applied shear strain rate of $1\,\text{sec}^{-1}$, and $t$ is time.

Fig. 2 shows the average shear stress-strain response for grains of different sizes where $\tau$ refers to the nominal (reaction) shear traction on the top surface. Conventional plasticity, on the other hand, has no explicit characterization of dislocations. In other words conventional plasticity may be recovered from PMFDM by setting $\boldsymbol{\alpha} = \boldsymbol{0}$ for all times and replacing (3) with

$$\boldsymbol{T} = \boldsymbol{C} : (grad\, \boldsymbol{u} - \boldsymbol{\varepsilon}^p) \; ; \; \dot{\boldsymbol{\varepsilon}}^p = \boldsymbol{L}^p$$
$$div\, \boldsymbol{T} = \boldsymbol{0}. \tag{17}$$

The average stress-strain profiles demonstrate that the smaller grains are indeed harder when compared to the larger grains, in accord with experimentally observed trends. The response of the $(100\mu m)^3$ sample is very close to the response as predicted from conventional plasticity.

It is observed that with the onset of plasticity the $\boldsymbol{\alpha}$ field becomes inhomogeneous. This is because the no-flow boundary condition imposed on dislocation density evolution induces a gradient in the $2-$direction in the $L_{21}^p$ component of $\boldsymbol{S}$ at the top and bottom boundaries, resulting in the development of an $\dot{\alpha}_{23}$ component, most likely with a gradient in the latter field. Low shear strain boundary layers are not observed and this is expected as a no-flow boundary condition (15) *does not* imply $S_{i2} = 0$ in the orthonormal basis $(\boldsymbol{e}_1, \boldsymbol{e}_2, \boldsymbol{e}_3)$ with $\boldsymbol{n} = \boldsymbol{e}_2$ for the top and bottom faces (Acharya and Roy, 2005). Fig. 3 provides some idea of the nature of inhomogeneous field profiles generated in this plastically constrained simple shearing deformation.

Though not very clear in Fig. 3, the $\alpha_{23}$ profile indicates high dislocation densities near the top and bottom boundaries in comparison to the grain interior. To better illustrate this feature, the following average

$$\tilde{\alpha}_{23}(x_2, x_3) = \frac{1}{a} \int_0^a \alpha_{23}(x_1, x_2, x_3) dx_1 \tag{18}$$

is plotted in Fig. 4, representing the variation of $\tilde{\alpha}_{23}$ along the $x_2$ direction at $x_3 = 0.5\mu m$ for $\Gamma = 0.2\%$ and $\Gamma = 0.8\%$. The dislocation density at the boundaries are much higher in magnitude in comparison to the grain interior. The profile displays a negative to positive and a positive to negative transition near the boundaries. The presence of a zero flow boundary condition on the surface causes sharp gradients in $\boldsymbol{\alpha}$ near the surface and the observed phenomenon could be a numerical artifact that is known to accompany the resolution of shocks. On the other hand, it is also possible that the presence of a sharp gradient in $\boldsymbol{\alpha}$ causes internal stress, and the observed overshoot/undershoot region may be physically representative of the development of a dipolar structure of dislocations to minimize this internal stress. At this time we are unable to strictly delineate the nature of the overshoot/ undershoot behavior.

The overall profile matches in sense with the 2-D discrete dislocation simulations as presented in Shu et al. (2001), suggesting that this feature is perhaps only related to the constrained boundary condition on plastic flow and the simple shearing deformation.

*4.1.2 Back stress and non-local effects in PMFDM*

An unloading step is performed for the $(1\mu m)^3$ block by applying kinematic boundary conditions similar to (16) on the nodes of the top, bottom, left and right faces with $\dot{\Gamma} = -1 \sec^{-1}$. The imposed displacements are such that $\Gamma = 0$ at the end of unloading. A strong Bauschinger effect is observed as shown in Fig. 2, purely due to non-local effects in the elastic strain as we show next.

Fig. 5 is a plot of the variation of various distortion rate components, volume averaged over the top layer of elements of the mesh. The mesh in question is a $32 \times 32 \times 1$ array of finite elements in the $x - y - z$ directions.

The notation $\langle * \rangle$ represents the volume average of the 12 component of the symmetric part of $*$ over the top layer of elements, and $\max_\bullet |*|$ represents the maximum of the absolute value of the $\bullet$ component(s) of $*$ over the top layer of elements.

The average $\langle * \rangle$ is relevant because the Bauschinger effect is read off from $\tau - \Gamma$ curves like Fig. 2 and Fig. 6, and it can be shown from the principle of virtual work, by choosing a test function $\delta u_2 \equiv \delta u_3 \equiv 0$ and $\delta u_1$ taking a value of unity on the top surface of the block and decreasing linearly to zero at the bottom of the top layer of elements, that the nominal surface traction $\tau$ is given by

$$\tau = \frac{1}{Ah} \int_{\substack{\text{top layer} \\ \text{of elements}}} T_{12} \, dv - \frac{1}{A} \int_{\substack{\text{left+right} \\ \text{faces of top layer} \\ \text{of elements}}} T_{11} \left\{ \frac{x_2}{h} - \frac{(H-h)}{h} \right\} da - \frac{1}{A} \int_{\substack{\text{front+back} \\ \text{faces of top layer} \\ \text{of elements}}} T_{13} \left\{ \frac{x_2}{h} - \frac{(H-h)}{h} \right\} da$$

, (19)

where $h$ is the element length in the $y$ direction and $A$ is the area of the top surface of the block (See Fig.1). Clearly, in the limit $h \to 0$ the last two terms in (19) vanish (provided the stress components are non-singular). Fig. 11 indicates that in our calculations, even at finite $h$ the contributions from the latter area averaged forces vanish. Incidentally, this particular test function can be represented exactly in our FEM calculation. Thus, isotropic elasticity and (19) imply that any deviations in the evolution of $\tau$ during unloading can be understood by monitoring averages of the $\langle * \rangle$ type, corresponding to various distortion rates that make up $\dot{U}^e$.

Fig. 6 indicates the range of applied strain along the unloading curve in Fig. 2 where we perform our analysis. The vertical lines indicate the range of applied strain where the unloading curve bends noticeably from the elastic unloading curve. The elastic unloading curve would be the result for the surface traction variation with $\Gamma$, under the specified displacement boundary conditions for simple shearing, in linear elasticity as well as conventional, rate-insensitive plasticity. The same strain range is also marked off by vertical lines in Fig. 5. Fig. 3 shows the average $\tau - \Gamma$ response along with field plots of $g, T_{12}, \varepsilon_{12}, |\alpha|$ and $\alpha_{23}$ profiles on the undeformed configuration, for the grain at $\Gamma = 0.72\%$. It is worth noticing that the state of the body is strongly inhomogeneous near the beginning of this unloading range.

Let us denote the time-dependent displacement field corresponding to the simple shearing homogeneous deformation satisfying the boundary conditions as $c$. The line representing $\langle grad\, \dot{c} \rangle$ during unloading is shown in Fig. 5. This line would also represent the variation of the top-layer-average of the elastic strain rate in linear elasticity as well as conventional, rate-insensitive, plasticity. A Bauschinger effect requires an averaged elastic strain rate curve above this line.

Next we consider $\langle grad\, \dot{z} \rangle$. Since $\max_{12,21} |S| \approx 0$, the particular form of the $\langle grad\, \dot{z} \rangle$ curve can only be attributed to non-local effects arising from $S \neq 0$ in the interior of the body, in particular from the regions of high negative shear stresses (w.r.t the rectangular Cartesian basis being utilized) as can be seen in Fig. 3.

The occurrence of the Bauschinger effect in our model is solely related to sources of internal elastic stress. For definiteness, we consider the case of all-round applied boundary conditions on the displacement field, which is also relevant to the simple shearing problem being considered in this example. Additionally, for the sake of this argument, we consider the fields $z$ and $\chi$ as known and the displacement boundary condition to be $u = \hat{u}$ on $\partial B$. Then the displacement field has to satisfy

$$div\{C:(grad(u-z)+\chi)\}=0, \qquad (20)$$

and, by the standard uniqueness theorem of linear elastostatics, can be written as the sum

$$u = z + u^z + u^\chi, \qquad (21)$$

where

$$\begin{aligned}
&div\{C:grad\, u^z\} = 0 \quad \text{on } B \\
&u^z = \hat{u} - z \quad \text{on } \partial B \\
&div\{C:grad\, u^\chi\} = -div\{C:\chi\} \quad \text{on } B \\
&u^\chi = 0 \quad \text{on } \partial B.
\end{aligned} \qquad (22)$$

Thus, stress in the body,

$$T = C:\{grad(u^z + u^\chi) + \chi\}, \qquad (23)$$

can only arise due to

1. a mismatch between the boundary values of the 'plastic displacement' field, $z$, and applied displacement boundary condition, $\hat{\boldsymbol{u}}$, and
2. the presence of a non-vanishing dislocation density field $\boldsymbol{\alpha}$ resulting in the incompatible elastic distortion field $\boldsymbol{\chi}$ that may cause stress by itself as well as serving as the source of an apparent body force in the solution for $\boldsymbol{u}^\chi$.

However, these two conditions above are not sufficient for stress in the body as non-vanishing symmetric parts are required of the fields $grad\,\boldsymbol{u}^z$, $grad\,\boldsymbol{u}^\chi$, and $\boldsymbol{\chi}$. By superposition, the displacement (and consequently the stress) due to the boundary-mismatch condition (21) above can also be viewed as the sum of two components, $\boldsymbol{u}^z = \boldsymbol{u}' + \boldsymbol{u}''$, with $\boldsymbol{u}'$ arising purely from satisfying the applied displacement condition and $\boldsymbol{u}''$ arising due to the boundary values of the plastic displacement field being different from zero.

Finally, we note that the equations (20)-(23) all hold in rate-form, i.e. if all the fields were to be replaced by their time-rates of change.

Reverting to the simple-shear problem and in light of the above, *if* the plastic displacement-rate vanished on the boundary $(\dot{z} = \boldsymbol{0}$ on $\partial B)$ and the incompatible elastic distortion rate were also to vanish $(\dot{\boldsymbol{\chi}} = \boldsymbol{0}$ on $B)$, i.e. the only possible source of stress was to be the applied boundary condition, equilibrium would demand that the symmetric part of $grad\,\dot{\boldsymbol{u}}$ be equal to $grad(\dot{\boldsymbol{c}} + \dot{\boldsymbol{z}})$, thus yielding $\left\langle grad\,\dot{\boldsymbol{U}}^e \right\rangle\big|_{\substack{\dot{\boldsymbol{\chi}} \equiv \boldsymbol{0} \text{ on } B \\ \dot{z} = \boldsymbol{0} \text{ on } \partial B}} = \left\langle grad\,\dot{\boldsymbol{c}} \right\rangle$ and the top-layer average of displacement gradient rate as $\left\langle grad\,\dot{\boldsymbol{u}} \right\rangle\big|_{\substack{\dot{\boldsymbol{\chi}} \equiv \boldsymbol{0} \text{ on } B \\ \dot{z} = \boldsymbol{0} \text{ on } \partial B}}$ shown in Fig. 5. Thus there would be no Bauschinger effect in our model under these conditions, but there would be inhomogeneous total deformation controlled by the field $z$.

Of course, the no-flow boundary condition imposed on dislocation density evolution results in the development of an inhomogeneous $\dot{\alpha}_{23}$ field, as discussed in Sec. 4.1.1. Thus, $\dot{\boldsymbol{\chi}}$ is non-zero in the body and there is no reason for the potential field $(\dot{z})$ of the compatible part of the slipping distortion to match the applied boundary values of the displacement rate field. These reasons result in the differences shown between $\left\langle grad\,\dot{\boldsymbol{u}} \right\rangle\big|_{\substack{\dot{\boldsymbol{\chi}} \equiv \boldsymbol{0} \text{ on } B \\ \dot{z} = \boldsymbol{0} \text{ on } \partial B}}$ and $\left\langle grad\,\dot{\boldsymbol{u}} \right\rangle$ in unloading shown in Fig. 5 as well as the difference between $\left\langle grad\,\dot{\boldsymbol{U}}^e \right\rangle\big|_{\substack{\dot{\boldsymbol{\chi}} \equiv \boldsymbol{0} \text{ on } B \\ \dot{z} = \boldsymbol{0} \text{ on } \partial B}}$ and $\left\langle grad\,\dot{\boldsymbol{U}}^e \right\rangle$. What we find somewhat remarkable is that there is no constitutive control on the direction in which the overall stress-strain curve should bend due to these internal stress effects and yet the bend does occur in the physically expected manner (for metallic materials, at least).

*Thus, the strong Bauschinger effect shown in Fig. 2 is completely a manifestation of internal elastic stress in* PMFDM. Interestingly, the deviation in $\langle grad\,\dot{\boldsymbol{u}}\rangle$ and $\langle grad\,\dot{\boldsymbol{U}}^e\rangle$ from the homogeneous solution arises in the presence of $\max_{\text{all }i,j}|\dot{\boldsymbol{\alpha}}|h \approx 0$ and $\max_{\text{all }i,j}\boldsymbol{S} \approx 0$, attesting to the spatial nonlocality inherent in these effects.

We note here that

1. The nonlocal internal stress effects do not introduce any material length-scale parameters in PMFDM (equations (1)-(2), where the constitutive equation for $V$ could very well not involve a physical length-scale, and yet there would be nonlocal effects). However, these equations do introduce a scale-dependence on the geometric size of the sample, as is revealed by dimensional analysis with the dislocation density treated as prescribed data[1].
2. The observed back stress in our model is not related to the phenomenological modeling of dislocation correlation effects motivated from coarse-graining of the dislocation evolution equation (Groma, 1997; Yefimov et al., 2004), or to additional configurational stresses arising from postulated dependencies of the free-energy on the Nye tensor (Menzel and Steinmann, 2000; Gurtin, 2002; Arsenlis et al., 2004).
3. Even though the Bauschinger effect in our model arises purely from internal elastic stress effects, it appears reasonable to ask as to what additional contribution to such should arise from the localized discreteness of dislocation cores. Careful comparisons of moving pile-up configurations within a 1-d, smeared continuously distributed setting and the discrete dislocation setting (Nadgornyi, 1988) show important differences. Such discreteness effects can be accounted for in a field setting like ours through a physically rigorous accounting of nonlinear crystal elasticity at the FDM level. The primary effect in PMFDM (via averaging, see Part I) would be through a different structure of the force equilibrium equation, apart from expected effects in the constitutive equations for $V$, $\boldsymbol{L}^p$.

### *4.1.3 Dependence of stress-strain response on sense of loading and initial dislocation density*

In this section we show that a dependence on the sense of loading can arise in PMFDM in the presence of an initial GND density distribution in the body. This effect arises due to the dependence of the GND velocity on the sign of the GND.

---

[1] It is a pleasure for AA to acknowledge a stimulating discussion with Marc Geers on this matter.

A $(1\mu m)^3$ constrained grain is considered as in the previous subsections. During the loading step, displacements corresponding to simple shearing is imposed on the nodes of left, right, top and bottom surfaces of the block as in (16). We refer to shearing in the positive $x_1$ direction as positive shear and in the negative $x_1$ direction as negative shear.

Here we study the following cases with an initially homogeneous $\alpha_{23}$ dislocation density distribution with $\alpha_{ij} = 0$ and $i \neq 2$ and $j \neq 3$.

Case 1: $\alpha_{23} = 4.05 \times 10^{-4}/\mu m$ applied positive shear.
Case 2: $\alpha_{23} = -4.05 \times 10^{-4}/\mu m$ applied positive shear.
Case 3: $\alpha_{23} = 4.05 \times 10^{-4}/\mu m$ applied negative shear.
Case 4: $\alpha_{23} = 0$ applied positive shear.
Case 5: $\alpha_{23} = 0$ applied negative shear.

Fig. 7 shows the average $\tau - \Gamma$ response for cases described above. When the grain is initially dislocation free then the stress-strain response is independent of the straining direction (Cases 4, 5). Interestingly, if the grain has an initially homogeneous, stress-free, edge dislocation density of $\alpha_{23} = 4.05 \times 10^{-4}/\mu m$ then the response is found to be straining direction dependent (Cases 1, 3), a feature that is exactly similar to that for the cases of homogeneous $\alpha_{23} = 4.05 \times 10^{-4}/\mu m$ (Case 1) and $\alpha_{23} = -4.05 \times 10^{-4}/\mu m$ (Case 2) but applied shearing in the positive direction. We note that such a dependence on the sense of loading is generally not seen in conventional plasticity for the initial conditions typically employed. We discuss the difference in response between Cases 1 and 2. Similar logic explains the difference in response for Cases 1 and 3 and the similarity, at least around initial yield, in response between Cases 2 and 3.

Assume that the strength $g$ of the material for the two cases is the same in a small, but finite, time-interval around initial yield. Only the shear components of the stress tensor $T$ are active, namely $T_{12}$ and $T_{21}$, due to the imposed simple shearing deformation. Note that even though the initial GND fields are different in sign, they are homogeneous on the body and consequently do not contribute to stress. Then from (8) the strain rate would be

$$\dot{\gamma}|_{Case\ 1} = \dot{\gamma}|_{Case\ 2}. \tag{24}$$

The averaged dislocation velocity tensor, $V$, and $L^p$ would therefore be

$$V|_{Case\ 1} = -V|_{Case\ 2},$$
$$L^p|_{Case\ 1} = L^p|_{Case\ 2} \tag{25}$$

from (9), (7), (6)$_2$ and (6)$_1$ respectively. We assume that for the small time interval under consideration, the stress field is homogeneous as well as the strength field so that $curl\ L^p = 0$. Under these circumstances, the top boundary of the block changes type, between inflow and outflow, for the two cases being considered, i.e. in the first case the uniform initial density moves inwards from the top boundary, convecting GND density consistent with the imposed zero-flow boundary condition leading to an inhomogeneous GND field, whereas in the second, the initial density moves towards the top boundary where it is confronted with the zero-flow boundary condition, leading to the development of a sharp gradient, at least. The two situations lead to different types of inhomogeneous GND fields which are most likely associated with different stress fields near the top boundary of the block. Consequently, we see the difference in stress strain response between Cases 1 and 2.

### *4.1.4 Convergence of results with respect to mesh refinement*

Figures 8, 9 and 10 display the variation of computed results with respect to mesh refinement. The convergence of results with respect to time-step refinement, for the time-step magnitudes used, has been verified separately.

Fig. 8 indicates a gradual increase in applied (reaction) traction with mesh refinement. It is important to expect a physically realistic upper bound to this increase. Fig. 11 demonstrates the plausibility of such a bound. Based on extrapolating the available data, a plausible, conservative upper bound for $\tau/\mu$ as $h \to 0$ appears to be $0.88 \times 10^{-3}$, which corresponds to a stress that is $\sim 5\%$ of the corresponding stress for a purely elastic material for the same level of applied strain. We expect this bound to not be violated as the increase in applied traction is related to the refinement and increase in magnitude of the GND profiles (Fig. 10), and an increase in the GND magnitude can only lead to the slowing down of this increase, due to a corresponding increase in the strength, $g$, which controls the magnitudes of both the GND velocity and the SSD slipping rate (Eqns. (4), (5), (6), (8), and (9)). We consider the trend displayed by the curves for $\langle g/\mu \rangle$ and $\langle |T'|/(\sqrt{2}\mu) \rangle$ for decreasing $h$ as indicative of this fact. Here, the notation $\langle * \rangle$ represents the average of $*$ over the top layer of elements. We also note that the difference of the result for the coarse mesh of $h = 0.05\mu m$ $(20 \times 20 \times 1)$ and the extrapolated upper bound is only $\sim 7\%$, and therefore the difference corresponding to the actual limit value as $h \to 0$ is expected to be lower if our hypothesis on the trend of the increase mentioned above holds.

Turning next to the question of convergence of the dislocation density for this specific problem, Figs. (9, 10) indicate a progressive refinement of the microstructure with decreasing mesh size, while coarser patterns seem to converge. To interpret the objectivity of such results, we consider various $L^p$ norms of the computed dislocation density fields for different mesh sizes:

$$\left|\boldsymbol{\alpha}_{23}^{h}\right|_{\infty} := \max_{\boldsymbol{x} \in B} \left|\boldsymbol{\alpha}_{23}^{h}\right|$$

$$\left|\boldsymbol{\alpha}^{h}\right|_{2} := \left(\int_{B} \left(\boldsymbol{\alpha}^{h} : \boldsymbol{\alpha}^{h}\right) dv\right)^{\frac{1}{2}} \quad (26)$$

$$\left|\boldsymbol{\alpha}^{h}\right|_{1} := \int_{B} \left(\boldsymbol{\alpha}^{h} : \boldsymbol{\alpha}^{h}\right)^{\frac{1}{2}} dv.$$

The max in $(26)_1$ makes sense as $\boldsymbol{\alpha}_{23}^{h}$ is a continuous function by definition of the adopted finite element basis functions. The choice of the tensorially non-invariant $L^{\infty}$ measure, $\left|\boldsymbol{\alpha}_{23}^{h}\right|_{\infty}$, is simply for the illustration of a particular result and bears in no way on the invariance of our theory or computation. The variations of these norms with $h$ are displayed in Fig. 12. The data are fitted by a best fit singular power function as well as a quadratic. In every case, the quadratic function yields a better fit in terms of residual errors in the fitting. Based on this fitting, *if* it were to be assumed that $\left|\boldsymbol{\alpha}^{h}\right|_{p}, p = \infty, 2$ is bounded as $h \to 0$, then a theorem of functional analysis (e.g. Evans, 1988) indicates that a subsequence of the $h$-parametrized family of functions $\boldsymbol{\alpha}^{h}$ converges *weakly* (weak *, for $p = \infty$) to a function in $L^p$, i.e. in the sense of all possible spatial weighted averages on the body corresponding to weighting functions in $L^q$, with $\frac{1}{p} + \frac{1}{q} = 1$. The theorem is actually valid for $1 < p \leq \infty$, with the case $p = 1$ leading to weak convergence of measures (Evans, 1988). Moreover, when weak convergence of a sequence can be established, the existence of a family of probability measures (Young measure) parametrized by position on the body is also guaranteed, and this helps in the probabilistic interpretation of the limit values of nonlinear functionals of the functions involved in the sequence. Similar ideas are applicable to $h, \Delta t$ parametrized families of approximate solutions. However, the actual construction of the limit probability measure may not be a straightforward proposition.

Interestingly, the above considerations lead to us to consider some natural, albeit heuristic, connections with modern PDE theory for conservation laws involving transport phenomena. Tartar (1979) pioneered the use of the Young measure limit of a sequence of functions as an ingredient in proving the convergence of the viscosity method for hyperbolic conservation laws. Subsequently, DiPerna (1985) defined the notion of a measure-valued solution to a conservation law and E and Kohn (1991), influenced by the work of D. Serre, showed that the physically relevant class of measure-valued solutions should be the ones that arise as limits of sequences of oscillatory, approximate, classical solutions so that uniqueness, with respect to initial data, may be expected.

These mathematically rigorous ideas with a very physically natural flavor lead us to consider the following practical use of our finite element simulations that generate approximate solutions subject to the physical balance laws. Consider an initial boundary value problem set up as we have been considering in this paper. Say one performs a sequence of finite element simulations for varying mesh size. At every (a select grid of) point(s) $(x,t)$ and for each basic field, record the values of the field obtained corresponding to the sequence of simulations. Construct a probability distribution, at each point $(x,t)$ and for each of the basic fields, from these observations in the natural way, starting from a frequency interpretation and improving on it by adapting ideas from the proof of existence of the Young measure as far as practically possible. Such a construction may be thought of as an analog of a measure-valued solution in the discrete context. In case the sequence converges at a specific $(x,t)$, the probability distribution at that $(x,t)$ should be sharply-peaked at the converged value. In problems where convergence of the sequence to a specific value at one or many points of space-time is found to be problematic, one may average the ingredients of the field equations with respect to these distributions to generate average fields and see how well these averaged fields satisfy the discretized field equations - if the determining solution sequence shows weak convergence, then the weak form of the finite element equations imply that such a test should be passed almost as a matter of definition. Thus, difficulties related to strong convergence of numerical results maybe rationally avoided when they are simply a result of the complicated physics involved and strong convergence is not even to be mathematically expected. Of course, such a methodology does not say anything about the correctness of the computational results so produced. However, the developed measures (probability distributions) may be compared against exact measure-valued solutions to the same problem, whenever the latter are available. In particular, note that if a classical solution is known to exist to a particular problem then it would correspond to a measure-valued solution that is a Dirac-delta function centered at an appropriate point in state space for each $(x,t)$. For a problem set-up that may be judged to provide universal results, the outlined procedure may be expected to deliver useful, mesh-objective results on the evolution of microstructural *fields* and their effect on mechanical properties. Moreover, in the case when the solution of a specific problem is expected to be classical (e.g. a spatially homogeneous solution), this philosophy can be executed smoothly (there is nothing to be done as one has strong convergence) as opposed to a probabilistic formulation where one solves PDE for the distribution functions themselves – practically speaking, resolving delta functions in high-dimensional states spaces, apart from the expense, is not a simple matter.

In concluding this section we note that the increasing microstructural refinement observed in our results is not due to the absence of physical length scales in the theory but due to the inherent oscillation and concentration properties of solutions of nonlinear transport processes (Tartar, 2002). Denoting by $\Gamma, \dot{\Gamma}, H$ the total applied engineering strain, strain rate and the dimension of the body, $\alpha_0$ as some representative measure of the magnitude of the initial dislocation density field and $x_i$ as generic coordinates on the body, dimensional analysis implies that the relations

$$\tau = \mu \Phi_1\left(\frac{\theta_0}{\mu}, \frac{g_s}{\mu}, \frac{g_0}{\mu}, \frac{\dot{\Gamma}}{\dot{\gamma}_0}, \frac{b}{H}, \alpha_0 H, m, \Gamma, k_0, \eta\right),$$

$$H|\boldsymbol{\alpha}| = \Phi\left(\frac{\theta_0}{\mu}, \frac{g_s}{\mu}, \frac{g_0}{\mu}, \frac{\dot{\Gamma}}{\dot{\gamma}_0}, \frac{b}{H}, \frac{x_1}{H}, \frac{x_2}{H}, \frac{x_3}{H}, \alpha_0 H, m, \Gamma, k_0, \eta\right), \quad (27)$$

$$b|\boldsymbol{\alpha}| = \tilde{\Phi}\left(\frac{\theta_0}{\mu}, \frac{g_s}{\mu}, \frac{g_0}{\mu}, \frac{\dot{\Gamma}}{\dot{\gamma}_0}, \frac{b}{H}, \frac{x_1}{H}, \frac{x_2}{H}, \frac{x_3}{H}, \alpha_0 H, m, \Gamma, k_0, \eta\right)$$

have to be satisfied, where $\Phi_1, \Phi, \tilde{\Phi}$ are functions of the arguments shown. The dimensionless arguments $b/H, \alpha_0 H$ introduce a dependence of the response on the geometric proportion of the body. It may also be expected that the physical scale of the microstructural refinement is affected by the nondimensional parameters $b/H, \alpha_0 H$.

## *4.2 A stability analysis of the time-dependent, spatially homogeneous, simple shearing solution in PMFDM and the possibility of including diffusion in GND evolution*

In this section we carry out numerical experiments related to the homogeneous, simple shearing solution in PMFDM. Blocks of different sizes, with dimensions and material properties similar to those used in Section 4.1, are sheared under displacement boundary conditions exactly as in Sec. 4.1.1. $32 \times 32 \times 1$ meshes are used for all blocks. For boundary conditions on the dislocation flow, $\boldsymbol{\alpha}(\boldsymbol{V} \cdot \boldsymbol{n})$ is set to zero on inflow boundaries at all times and $\boldsymbol{L}^p$ on the boundary is set to the value obtained from the simple shearing solution for conventional plasticity governed by Voce law hardening. The initial condition on dislocation density is set to zero. Under these conditions, it is easily verified that the time-dependent, homogeneous, simple shearing solution of conventional plasticity is also a solution in PMFDM, independent of the geometric size of the block.

Fig. 13 shows the average stress-strain response for grains of different sizes. The average stress-strain responses for grain sizes smaller than approximately $(1\mu m)^3$ eventually deviate from the conventional plasticity response, demonstrating a harder trend. We have determined that for the larger grains, this phenomenon is observed at a much higher average strain level, if at all. For example in the $(1\mu m)^3$ grain the deviation is noticed at approximately 2% strain. On close analysis we observe that the deviation from conventional plasticity response for the $(0.5\mu m)^3$ grain is due to a gradual increase in the dislocation density that eventually increases the overall strength in the grain. Such an increase in strength below the 1μm threshold has interesting implications in terms of possible connections between physical instability predicted by theory and experimental observations of drastically different behavior, with respect to size effects, of free standing polycrystalline gold films below 1µm thickness under nominally homogeneous deformation (Espinosa et al., 2004), as compared to films of thickness above 1µm. Of course, given the novelty of our computational undertaking, the flip side of the expectation, i.e. some numerical difficulty masquerading as a prediction of physical instability, has to be considered.

Without further nonlinear analysis, we surmise that there are two possible reasons for such a deviation. First, without rigorous numerical analysis of our computational scheme, the possibility of numerical inaccuracies cannot be ruled out. However, we use consistent and stable schemes that have been tested on corresponding linear problems. If anything, the GLS method that we use for the dislocation density evolution actually has a fairly heavy dose of consistent diffusion built into it so that it is unlikely that round-off errors can grow unchecked in the absence of physical instability implied by the theoretical model. In fact, given the fixed meshes for all problems and the apparent instability we observe with decreasing size of specimen, the result for the $(1\mu m)^3$ block may be assumed to be a conservative convergence check for the results for the $(10\mu m)^3$ and $(100\mu m)^3$ blocks.

This leads us to speculate on a second possible reason for the behavior of the solution for the $(0.5\mu m)^3$ sample – if it were to be proven that the time-dependent, spatially homogeneous simple shearing solution of the theory was unstable, with the instability threshold depending inversely on grain size and directly on applied strain magnitude, then the observed result in Fig. 13 may not be considered as disconcerting. We are of the opinion that numerical computations with reliable computational schemes are a good test for probing such theoretical instability – numerical round-off may be otherwise interpreted as good sources of random perturbations; for a physically stable solution solved with an adequate numerical scheme, such perturbations do not grow. On the other hand, if a base solution is physically unstable according to the equations of the model, then the round-off-perturbed computational solution can diverge (without finite-time blow-up) from the base solution. Given the nonlinear transport nature of PMFDM, such controlled instabilities need not be surprising. Since the primary conclusion here is related to proving instability, a nonlinear study of the stability of the time-dependent homogeneous solution is warranted, with linear stability analysis likely to play a limited role. However, a rough linear analysis of a model appears to support our numerical observations as we show next.

### 4.2.1 *Boundedness of solutions of a linear Cauchy problem with respect to initial data*

We analyze a model problem in 1-d that reflects the essentials of the coupling between the dislocation density $\alpha$ and the strength $g$. Consider the system of nonlinear equations

$$\dot{g} = \rho_0(g,\alpha)|\alpha|$$
$$\dot{\alpha} = \frac{\partial}{\partial x}\left(\rho_1(g,\alpha) + \rho_2(g)\right), \tag{28}$$

where $\rho_0, \rho_1, \rho_2$ are prescribed functions. The function $\rho_1$ is such that $\frac{\partial \rho_1}{\partial \alpha}(g,0) = 0$ and $\rho_0 > 0$. The principal part of the linearized version of (28) about a spatially homogeneous base state $(\bar{g}, \bar{\alpha})$ may be written as the system

$$\frac{\partial}{\partial t}r = A\frac{\partial}{\partial x}r; \quad A = \begin{bmatrix} 0 & 0 \\ k_2 & k_1 \end{bmatrix}; \quad r = \begin{Bmatrix} \tilde{g} \\ \tilde{\alpha} \end{Bmatrix}, \tag{29}$$

with $k_1 = 0$ if $\bar{\alpha} = 0$ and $\tilde{g}, \tilde{\alpha}$ are perturbations.

The matrix $A$ has (real) eigenvalues $\lambda = 0$ and $\lambda^* = k_1$ with corresponding eigenvectors $\mu = \{-k_2 \quad k_1\}^T$ and $\eta = \{0 \quad 1\}^T$, respectively (the symbols $\mu, \eta$ of this subsection are not to be confused with the material parameters of the same name in the rest of the paper). When $k_1 = 0$, the eigenvalue $\lambda$ has multiplicity 2 with the only eigenvector $\eta$. Thus, in this case, $A$ does not have a full set of eigenvectors.

Physically, the case $k_1 = 0$ corresponds to a base state representing spatially homogeneous plastic strain distribution in PMFDM or the case of Lower-order Gradient Plasticity (LOGP).

We now consider the base state to be independent of time. Then, following standard ideas of Fourier analysis for linear constant coefficient problems (e.g. Kreiss and Lorenz, 1989), it suffices to consider initial data for (29) of the type

$$r(x,0) = e^{i\omega x}\hat{r}(\omega), \tag{30}$$

where is $\omega$ is an arbitrarily fixed wavenumber for perturbations, and $\hat{r}$ is a bounded function of $\omega$ as $\omega \to \pm\infty$. One now seeks solutions of (29)-(30) of the type $r(x,t) = e^{i\omega x}f(\omega,t)$, reducing the problem to the solution of a linear, constant-coefficient system of ODE.

Case 1: $k_1 \neq 0$. In this case, a solution to (29), (30) is

$$r(x,t) = c_1(\omega)e^{i\omega(x+\lambda t)}\mu + c_2(\omega)e^{i\omega(x+\lambda^* t)}\eta; \quad i := \sqrt{-1}$$
$$\begin{Bmatrix} c_1 \\ c_2 \end{Bmatrix} = [\mu \quad \eta]^{-1}\begin{Bmatrix} \hat{r}_1(\omega) \\ \hat{r}_2(\omega) \end{Bmatrix}. \tag{31}$$

Clearly, solutions are bounded with respect to $\omega$. Thus, in the presence of transport of $\alpha$, the linearized-constant-coefficient problem is strongly hyperbolic, even though it is not symmetric hyperbolic.

Case 2: $k_1 = 0$. Due to the unavailability of a full set of eigenvectors in this case, an extra time-dependent mode enters the solution (e.g. Boyce and DiPrima, 1977):

$$r(x,t) = c_1(\omega) e^{i\omega(x+\lambda t)} \eta + c_2(\omega) e^{(i\omega x + \bar{\lambda} t)} \{\bar{\eta} t + \xi\} \tag{32}$$

with $\lambda$ (scalar), $\eta$ (vector) as before and any choices $\bar{\eta}, \xi$ (vectors) and $\bar{\lambda}$ (scalar) satisfying

$$\begin{aligned}(i\omega A - \bar{\lambda} I)\bar{\eta} &= 0 \\ (i\omega A - \bar{\lambda} I)\xi &= \bar{\eta},\end{aligned} \tag{33}$$

where $I$ is the $2 \times 2$ identity matrix. $c_1(\omega), c_2(\omega)$ are determined as in $(31)_2$, but now involving the vectors $\eta, \xi$.

Motivated by the eigenpair $(0, \eta)$ of $A$, the choices

$$\begin{aligned}\bar{\lambda} &= 0 \\ \bar{\eta} &= i\omega\eta \\ \xi &= \{1/k_2 \quad 0\}^T\end{aligned} \tag{34}$$

satisfy (33). Thus, the solution (32) takes the form

$$r(x,t) = c_1(\omega) e^{i\omega x} \eta + c_2(\omega) e^{i\omega x} [i\omega t \eta + \xi]. \tag{35}$$

The solution in this case is not bounded independently of $\omega$ (due to the presence of the term $i\omega t\eta$) regardless of $t > 0$, however small. Hence, the linearized, constant-coefficient problem, even though weakly hyperbolic due to the presence of real eigenvalues, is 'ill-posed' according to conventional definitions, or weakly well-posed due to the linear growth, as opposed to exponential growth, in $|\omega|$. As a separate matter, all wavenumbers grow with time, with the highest rate of growth for the smallest wavelengths.

Notice that due to the form of $\eta$, the growth in solutions of (29) occurs in $\alpha$. Roughly speaking, in the context of PMFDM from a base state representing a conventional plasticity homogeneous solution, this would mean the development of a non-zero $\alpha$ solution. But then the linearized problem from a subsequent base state becomes strongly hyperbolic. Also, we expect that growth in $\alpha$ would raise the strength due to the form of (28) and this would cut off the plasticity embodied in the function $\rho_2$ in the actual nonlinear problem. The latter conclusion also applies to solutions of LOGP where the regularizing effect of $k_1 \neq 0$ (transport) is not available in the linearized model problem. For LOGP, this provides an interesting inhomogeneity-generating mechanism at small strains that may be of some use in the prediction of coarse-slip microstructures in conjunction with physically-motivated hardening rules (Asaro, 1983; Bassani, 1994).

Hence we conclude that it is natural to expect controlled growth of perturbations from a spatially homogeneous conventional plasticity solution within our model. In this connection, we have also verified that if the term $L^p$ is hard-coded (i.e. no perturbations allowed in this term) to be the time-dependent homogeneous solution of conventional plasticity, then the classical homogeneous solution is recovered in our numerical calculations described in this section, independent of specimen size. This fact is consistent with the weak hyperbolicity predicted in the linearized model problem due to the presence of the term $\rho_2$ in (28). With regard to the overall analysis, however, we note that it is silent about the dependence of the growth of perturbations on geometric size and level of applied deformation.

*4.2.2 Inclusion of a diffusive term in Dislocation Density evolution*

Strictly mathematically speaking, a diffusive term can be easily inserted into PMFDM that can eliminate the instabilities under discussion, if so desired. Since the term $\overline{\alpha \times V} - \overline{\alpha} \times \overline{V}$ is modeled phenomenologically (see Part I for the meaning of the overhead bars), if one assumes the constitutive equation

$$\overline{\alpha \times V} := L^p + \overline{\alpha} \times \overline{V} + \varepsilon \, curl \, \overline{\alpha} \tag{36}$$

where $\varepsilon$ is a strictly positive, scalar material parameter with physical dimension of $\left(length^2 / time\right)$, then (4) assumes the form of a diffusive transport equation

$$\dot{\alpha} = \varepsilon \, div(grad\,\alpha) - curl\left(L^p + \alpha \times V\right) + s, \tag{37}$$

where the constraint $div\,\overline{\alpha} = 0$ has been used. Indeed, on repeating the experiments of this section with (37) we observe no deviation from conventional plasticity response even at higher strains for all grain sizes, thus indicating that the homogeneous simple shearing solution appears to become stable.

However, for the overall theory, this is at the expense of physically ambiguous phenomenology needing an extra material parameter (or variable) definition and additional boundary conditions. Additionally, it may very well be that the original equation without the addition of diffusion is more representative of the transport of dislocations leading to mesoscopic plasticity as well as being far richer mathematically.

*4.3 Size Effects in an Infinitely Long beam*

Size effects are investigated on two infinitely long beams whose other spatial dimensions, namely the $width \times height$, are $1\mu m \times 1\mu m$ and $0.5\mu m \times 0.5\mu m$. A unit cell of $1\mu m$ length is analyzed for the thicker beam. For the thinner beam the unit cell length is $0.5\mu m$. In Fig. 1, $a = H = c = 1\mu m$ for the thicker beam and $a = H = c = 0.5\mu m$ for the thinner beam. A regular mesh of $32 \times 32 \times 1$ elements is used to discretize the two cubes so obtained.

The initial conditions imposed are as in Sec. 2. The imposed boundary conditions are as follows: The displacements on the bottom face are constrained in all three directions while those on the top, left and right faces are constrained in the $x_2, x_3$ directions only. The front and back faces are traction free. Displacements corresponding to a simple shear strain of 0.8% are prescribed on the top face through the kinematic boundary condition

$$u_1(t) = H\dot{\Gamma}t, \tag{38}$$

where $H$ is the height of the beam, $\dot{\Gamma} = 1 \text{ sec}^{-1}$ is the applied shear strain rate and $t$ is time. Periodic boundary conditions are imposed on the right and left faces as

$$u_1 \text{ on the right face}(x_1 = w) \;=\; u_1 \text{ on the left face}(x_1 = 0) \tag{39}$$

where $w$ is the representative unit cell length. The front, back, top and bottom faces of the beam are assumed to be plastically rigid and hence the zero-surface flow boundary condition (15) is imposed. Outflow of dislocations are allowed on the right and left faces and for this we apply the least restrictive boundary condition

$$\boldsymbol{\alpha}(\boldsymbol{V} \cdot \boldsymbol{n}) = \boldsymbol{0} \text{ on } \partial B_i \text{ and}$$

$$\boldsymbol{L}^P \times \boldsymbol{n} = \dot{\gamma}^t \frac{\left(\boldsymbol{T}'\right)^t}{\left|\boldsymbol{T}'\right|^t} \times \boldsymbol{n} \text{ on the surface,} \tag{40}$$

where $\partial B_i$ is the set of inflow points of the boundary (boundary points where $\boldsymbol{V} \cdot \boldsymbol{n} < 0$). As defined in Sec. 3, $(\bullet)^t$ refers to the value of $(\bullet)$ at the time $t$, and fields without a superscript refers to values at the current time $t + \Delta t$.

Fig. 14(a) is the average stress-strain response of the beams in simple shear. The thinner beam shows a harder response in comparison to the thicker beam, though the response is softer when compared to the average stress-strain response in Fig. 2. This is attributed to the fact that the plastic constraint from the right and left faces is significantly less in this case than its constrained grain counterpart which induces a lower dislocation evolution through (4). Also the free exit of GNDs at the outflow boundary points brings down the net dislocation density in the film. Fig. 14(b),(c),(d) are the inhomogeneous $\alpha_{23}, T_{12}$ and $\varepsilon_{12}$ profiles on the undeformed configuration for the 1μm thick beam at $x_3 = 0.5\mu m$, which are considerably different from the profiles for a constrained grain in Fig. 3. A no-flow boundary condition on the top and bottom surfaces does not pose any constraint on the $S_{i2}$ components of the slipping distortion tensor as discussed in Sec. 4.1.1. Interestingly, for this case low shear strain at the top and bottom boundaries are observed, as shown in Fig. 14(d). Also, the strain profile in Fig. 14(d) is inhomogeneous with a lack of symmetry[2]. However at a low strain of $\Gamma = 0.165\%$ the profile is more or less symmetric as shown in Fig. 15. We are inclined to attribute the loss of symmetry to an inherent sensitivity of our equations to perturbations. Random perturbations due to numerical round off affect the solution but gross features (stress-strain response) are reproducible.

### *4.4 Stressing of thin films due to thermal mismatch*

Though thin films in electronic circuitry do not serve a load bearing structural role, they are invariably subjected to high levels of mechanical stress as a result of constraints imposed by the material on which they are deposited.

Films deposited on a substrate are often coated with a thin oxide layer of low thermal coefficient of expansion, which is formed by self-passivation as in the case of Aluminum (Shen and Ramamurty, 2003; Kraft et al., 2002). The presence of a capping layer affects the response of the film to thermal stressing. Typically, the metallic film is deposited on a substrate at an elevated temperature. As the film cools, high residual stresses occur in the film from an initial stress-free state as a consequence of mismatch in the coefficients of thermal expansion of the substrate, film and the passivation layer (when present).

We consider a thin film with a thickness of 1μm. A unit cell of length 4μm and width 1μm is analyzed. A regular mesh of $20 \times 10 \times 1$ elements is used to discretize the unit cell. In Fig. 1 $a = 4$μm and $H = c = 1$μm.

The imposed initial conditions are as in Section 2. The imposed boundary conditions are as follows: For the right and left faces anti-periodic boundary conditions are imposed such that

$$u_1 \text{ on the right face}(x_1 = 4\mu m) = -u_1 \text{ on the left face}(x_1 = 0). \tag{41}$$

The substrate is assumed to act as a rigid constraint and hence displacements on the bottom face are constrained in all three directions. The front and back faces are traction free. To model the effect of a capping layer which is almost thermally inert, we constrain the $x_1, x_3$ displacement on the top face of the film as an approximation. The bottom face of the film being plastically rigid, a zero-flow boundary condition is appropriate. The same assumption may be made in the presence of a capping layer. Thus boundary condition (15) is imposed on the bottom face and the top face when the capping layer is present. For all the other faces we allow outflow of dislocations by applying the least restrictive boundary condition (40). The heating/cooling rates for the films are maintained at $10°$C/min. The reference strain rate, in (8), used for this numerical experiment is $\dot{\gamma}_0 = 9 \times 10°$C/min $= 3.92 \times 10^{-6}$ sec$^{-1}$.

---

[2] We thank J.C. Nagtegaal for noticing and pointing out this fact to us.

Thermal cycling of the films is initiated from an initial stress-free state at $600°C$. The films are then cooled to room temperature ($25°C$), heated to $500°C$, and finally cooled to $300°C$.

Fig. 16 is a plot of the volume average axial stress, $[T_{11}]$, versus temperature. The stress in the passivated film is observed to be higher than in the unpassivated film. The response when compared to that of conventional plasticity is significantly different as well. Bauschinger effect in both the capped and uncapped samples is observed. It is to be noted that even though we have not fitted any of the material parameters to that of the material used in Kraft et al. (2002), yet there are similarities in the obtained results. For example, the average axial stress for the capped film is higher than its uncapped counterpart both at room temperature and at $500°C$, with the capped film displaying an overall harder response. The thermal softening in heating observed in Kraft et al. (2002), is however not observed in our numerical experiments. Though we do not pursue this any further, this feature can be modeled by allowing the saturation stress ($g_s$) to be temperature dependent through a constitutive relation. Fig. 17 is a plot of the $T_{11}$ profile for the 1μm thick capped and uncapped film at $25°C$ on the undeformed mesh. The presence of the cap on the film affects the overall stress profile as demonstrated. This is expected, as the no-flow boundary condition imposed on the plastically constrained cap surface induces a growth in the net dislocation density and consequently the overall stress. For the uncapped film the free exit of GNDs at the outflow boundary points on the top surface brings down the net dislocation density, which explains the softer $[T_{11}]$ response of the uncapped film in Fig. 16.

Size effects in films under thermo-mechanical stresses are experimentally observed (Shen and Ramamurty, 2003). To investigate such effects, we consider an uncapped film with a thickness of 0.5μm. A unit cell of length 4μm and width 1μm is analyzed. All boundary conditions are similar to the unit cell considered for the uncapped thicker film analyzed before. In Fig. 1, $a = 4$μm, $H = 0.5$μm and $c = 1$μm. A regular mesh of $32 \times 10 \times 1$ elements is used to discretize the body. The film is cooled from an initial stress-free, dislocation-free state at $600°C$ to $250°C$ at a rate of $10°C/min$. The thinner film of 0.5μm thickness is harder than the thicker film, as demonstrated in Fig. 16, in accord with experimental results (Shen and Ramamurty, 2003). Figs. 18(a),(b) represent $T_{11}$ and $|\alpha|b$ profiles for the 0.5μm and 1μm thick uncapped films at $260°C$. The $x_2$ coordinate is scaled by the thickness $H$. Note that there is a lack of self-similarity between the plots for the two thicknesses in each case. The $T_{11}$ profile in the thinner film is noticeably higher, which explains the higher $[T_{11}]$ response for the thinner film in Fig. 16. The $|\alpha|b$ profile indicates that the average dislocation density is higher in the thinner film. The films demonstrate the formation of a boundary layer of high dislocation density content at the film-substrate interface whose width scales inversely with film thickness in the scaled coordinate in the film-thickness direction. This scaling of the boundary layer width with film thickness is similar to that observed in the 2D discrete dislocation simulations of Nicola et al., 2001. In Fig. 18(c) the $|\alpha|H$ profiles in the two films under consideration demonstrate (approximate) self-similarity. The absence and presence of self-similarity in the $|\alpha|b$ and $|\alpha|H$ profiles indicate the presence of a scaling of the form $b/H$ in $|\alpha|$ from (27)$_{2,3}$, with a weak dependence of the function $\Phi$ on $b/H$.

In concluding this section we note that the numerical results indicate that the imposed boundary conditions are less than perfect for the modeling of a unit cell representative of an infinite layer. We believe this is related to the imposed inflow-outflow boundary conditions on the dislocation flow. However, the results demonstrated are not expected to be substantially affected by this modeling flaw.

## 5. Concluding Remarks

A finite element based approximation tool for PMFDM has been shown to perform adequately for the modeling of mesoscopic plasticity. In particular, most of the commonly accepted physical benchmark problems in this new modeling arena have been tackled successfully. There are at least three ways in which significant size effects at initial yield are possible within our model on theoretical grounds. We are exploring these possibilities at the present time.

The nature of approximate solutions to our theory raises interesting questions related to connections with modern concepts of nonlinear PDE theory. Overall, the results reported in this paper appear to indicate that the developed tool can serve as a practical option for the modeling of micron scale plasticity. A particularly noteworthy feature, due to the phenomenological description of dislocation velocity involved, is the ability to conduct simulations at physically reasonable strain rates, while recovering essential physical predictions not dependent on the precise nature of the phenomenological assumption. As a rough estimate, strain rates for discrete dislocation simulations are typically of the order of $10^3 \text{sec}^{-1}$ or higher. On the other hand our numerical experiments are conducted at a strain rate of $1\text{sec}^{-1}$ and may be adjusted even further by a suitable choice of the material parameter $\dot{\gamma}_0$ (as in the thermal cycling examples of this paper). In our model for a $(1\mu m)^3$ grain under simple shearing up to $0.8\%$ strain (as in Sec. 4.1.1), the wall clock time for the loading step was $\sim 28$ hours $(\sim 1 \text{ day})^3$. Even if one allows a most favorable estimate for a discrete dislocation simulation for the same problem of only $\sim 1$ hour at a strain rate of $10^3 \text{sec}^{-1}$, a simulation carried out at a strain rate of $1\text{sec}^{-1}$ would involve a wall clock time of $\sim 41 \text{days}$.

---

[3] We have not paid attention to optimizing the numerical implementation reported here with regard to computational efficiency. In particular, avenues for parallel computation have not been pursued.

## 6. Acknowledgments

We are very grateful to Satya Varadhan and Armand Beaudoin for their kind help in getting AR started with the use of PETSc. Support for the work from the Program in Computational Mechanics of the U.S. Office of Naval Research, Grant No. N00014-02-1-0194 is gratefully acknowledged.

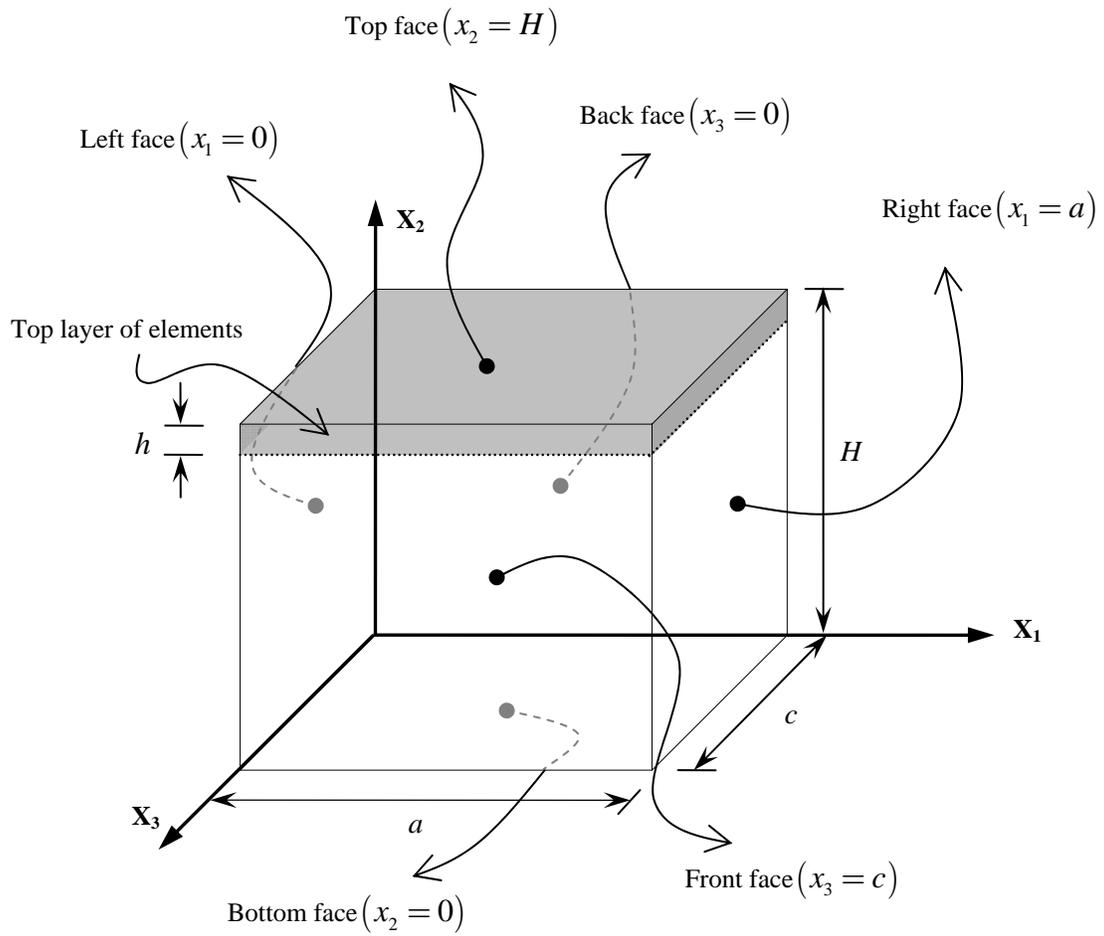

Fig. 1. Schematic layout of a typical model geometry.

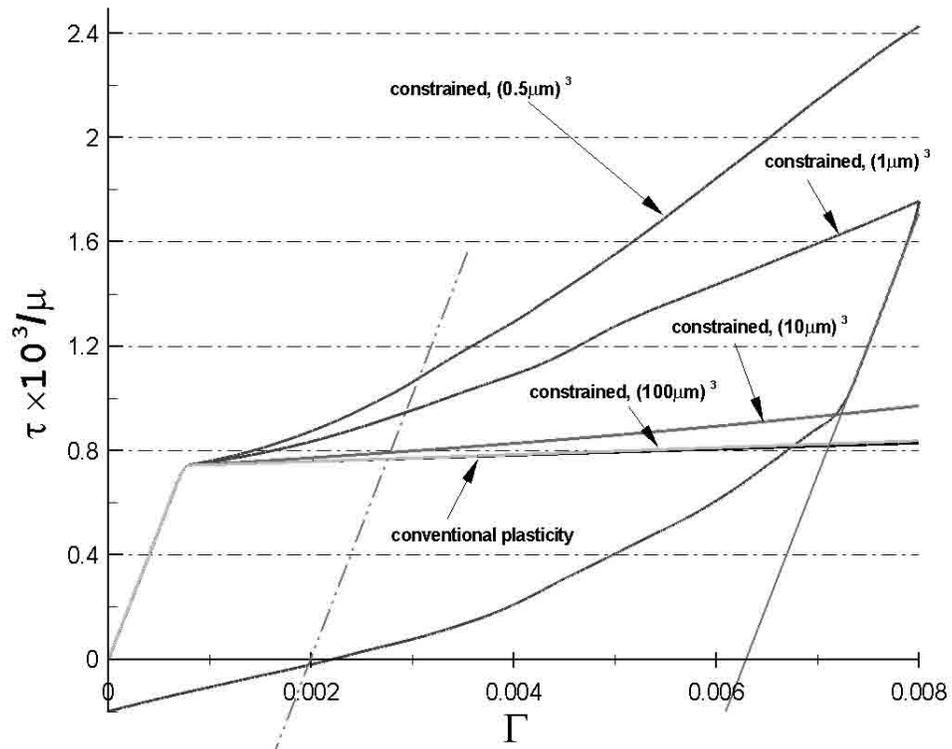

Fig. 2. Size effect on constrained grains in simple shear; Bauschinger effect in unloading.

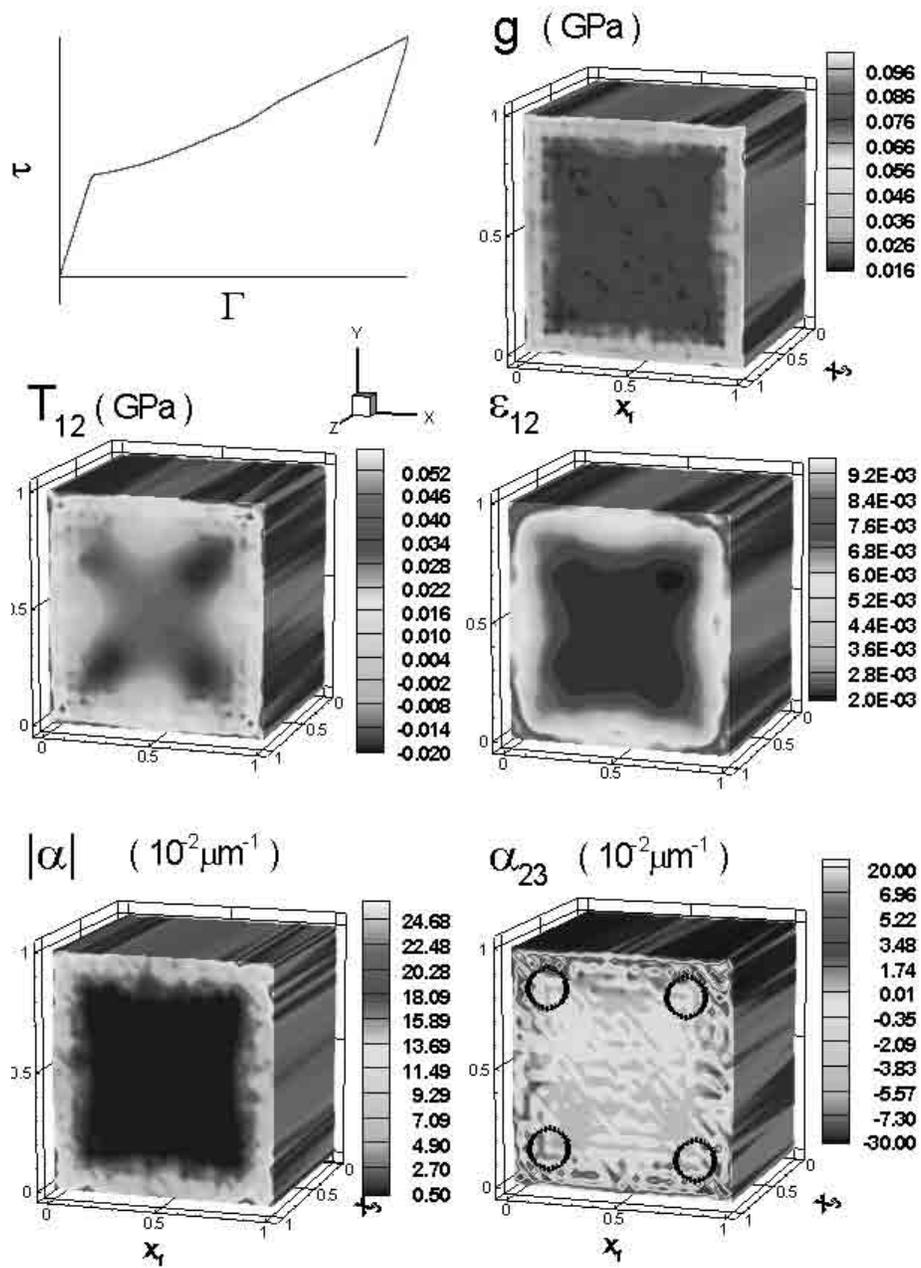

Fig. 3. Simple shear of a constrained grain; $g, T_{12}, \varepsilon_{12}, |\alpha|, \alpha_{23}$ on the undeformed mesh in unloading. Black circles in $\alpha_{23}$ mark regions of noticeable evolution near yield in reverse laoding. Spatial dimensions in $\mu m$.

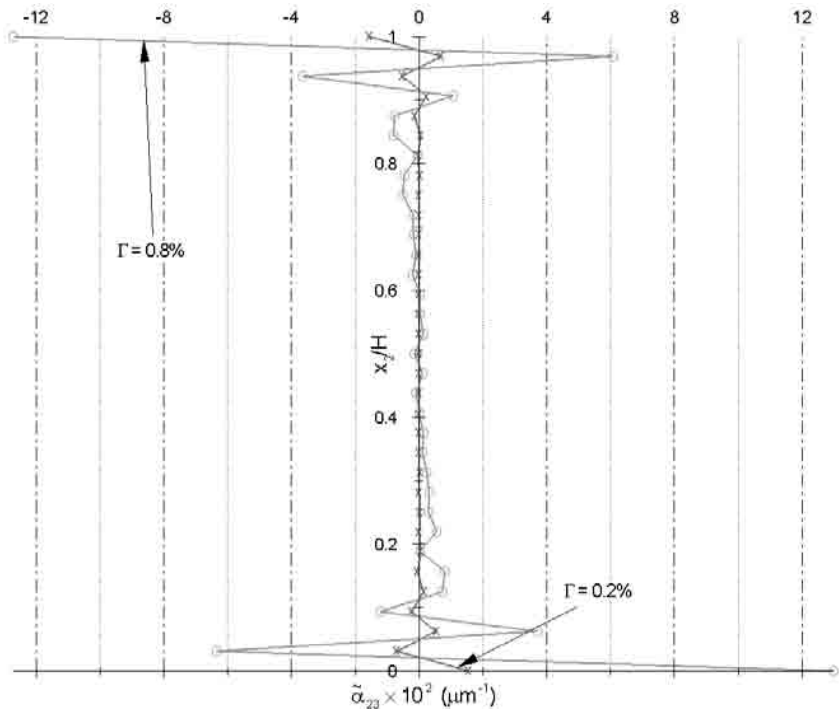

Fig. 4. Variation of $\tilde{\alpha}_{23}$ along $\dfrac{x_2}{H}$ at $x_3 = 0.5\mu m$.

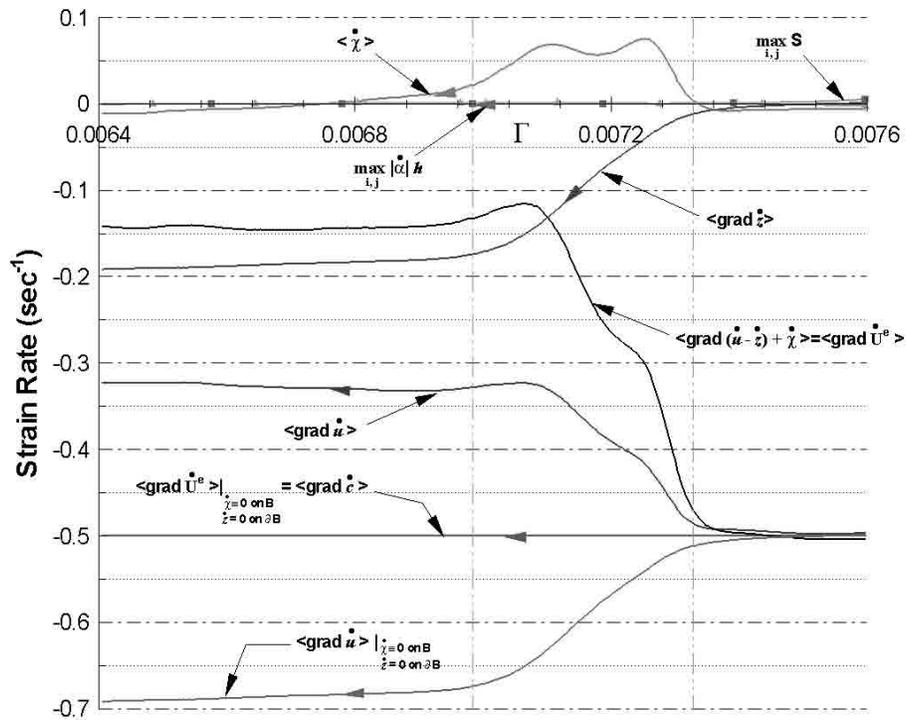

Fig. 5. Variation of rates of various distortion components with applied strain. The notation $\langle * \rangle$ represents the volume average of the $12$ component of the symmetric part of $*$ over the top layer of elements

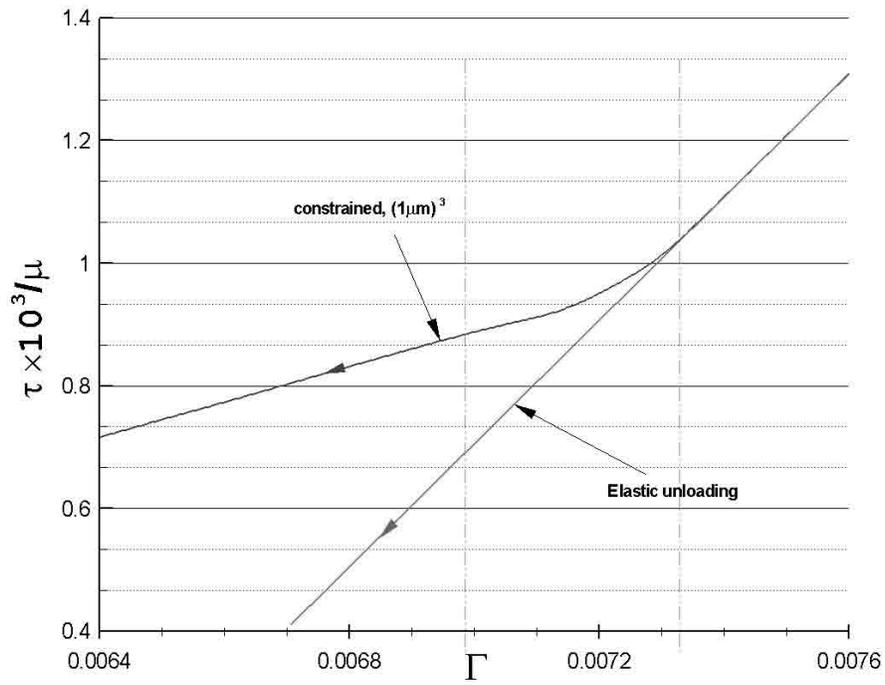

Fig. 6. Average $\tau - \Gamma$ response in unloading.

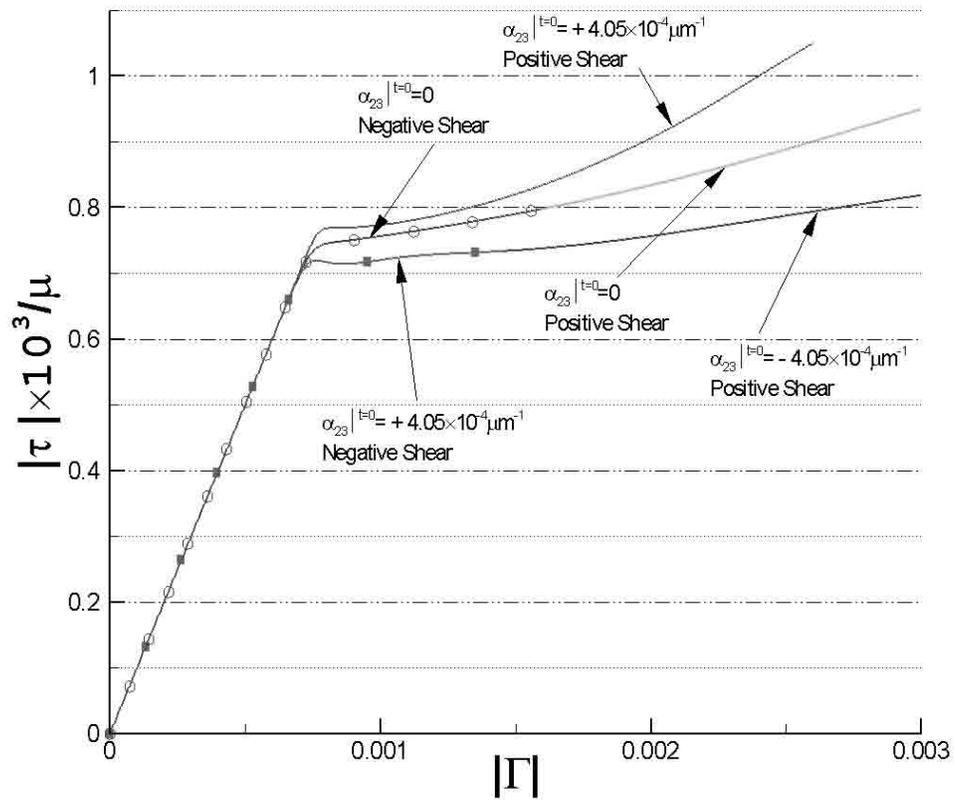

Fig. 7. Average $\tau - \Gamma$ response of a $(1\mu m)^3$ grain with different initial dislocation density and loading direction.

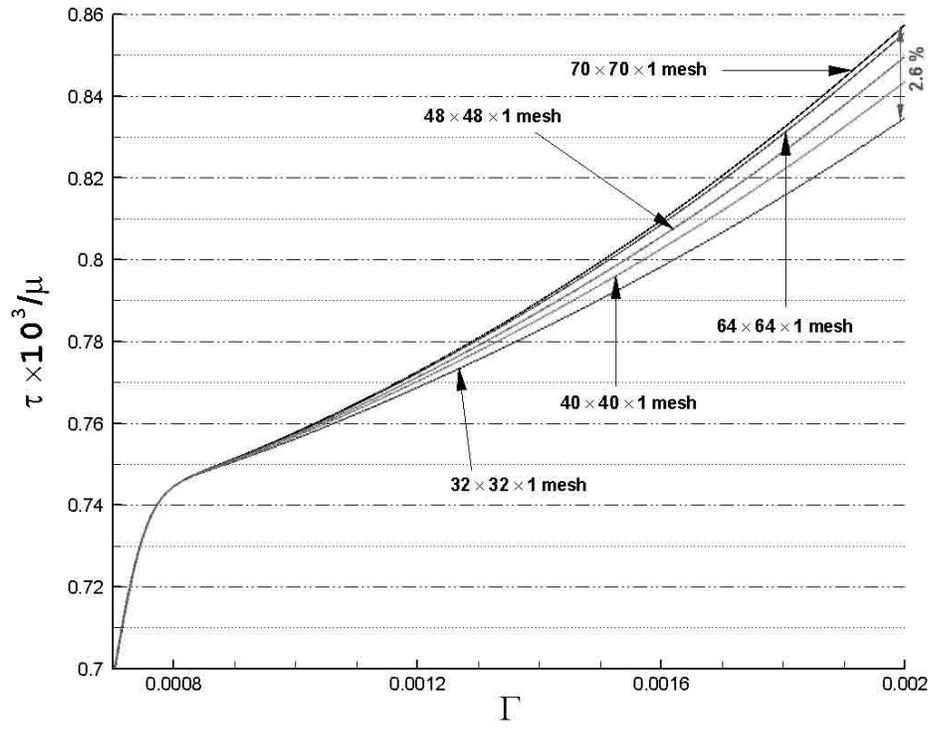

Fig. 8. Average $\tau - \Gamma$ response of a $(1\mu m)^3$ grain under simple shear for different mesh sizes.

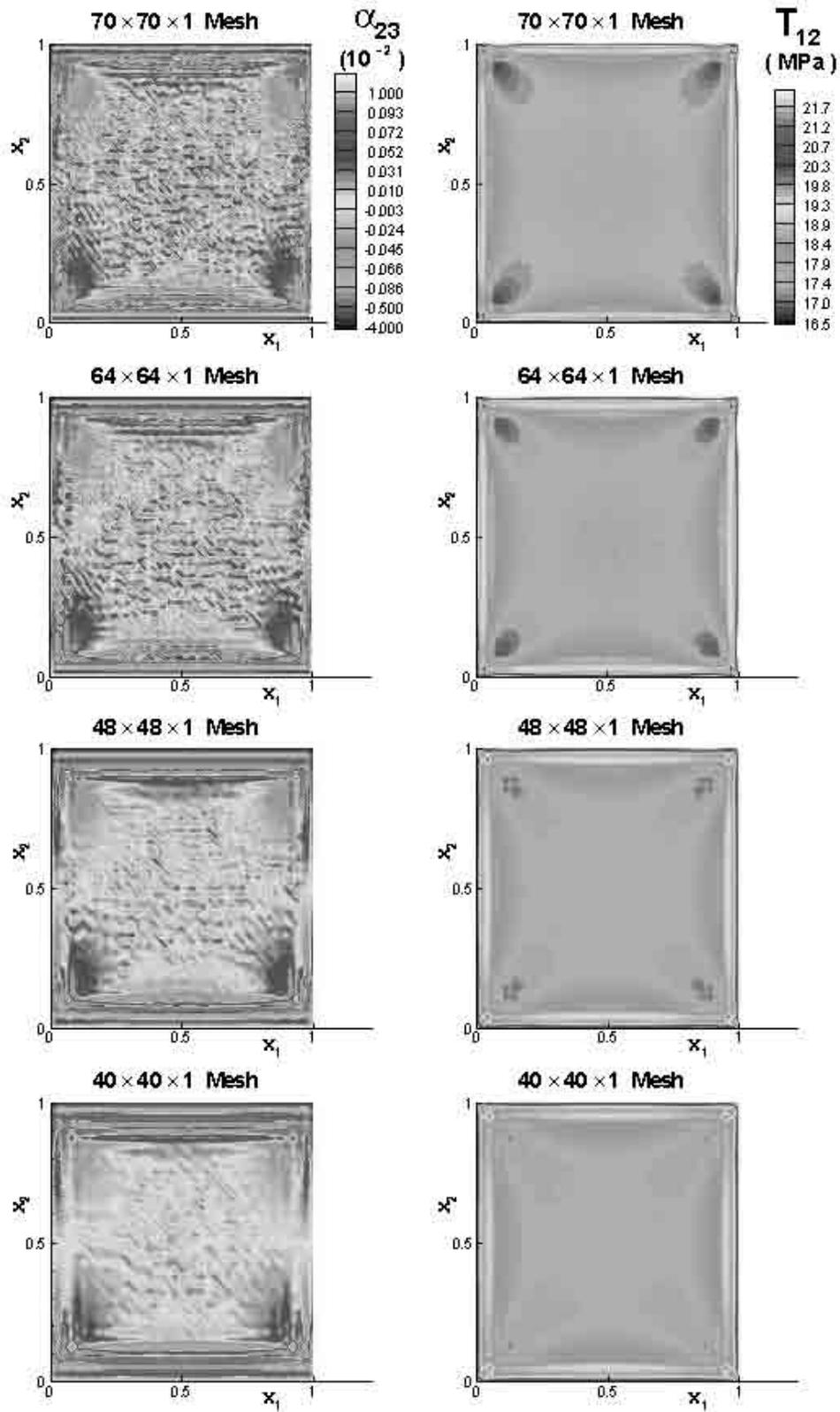

Fig. 9. $\alpha_{23}, T_{12}$ profile for a $(1\mu m)^3$ grain at $x_3 = 0.5\mu m$ for different mesh sizes at $\Gamma = 0.2\%$.

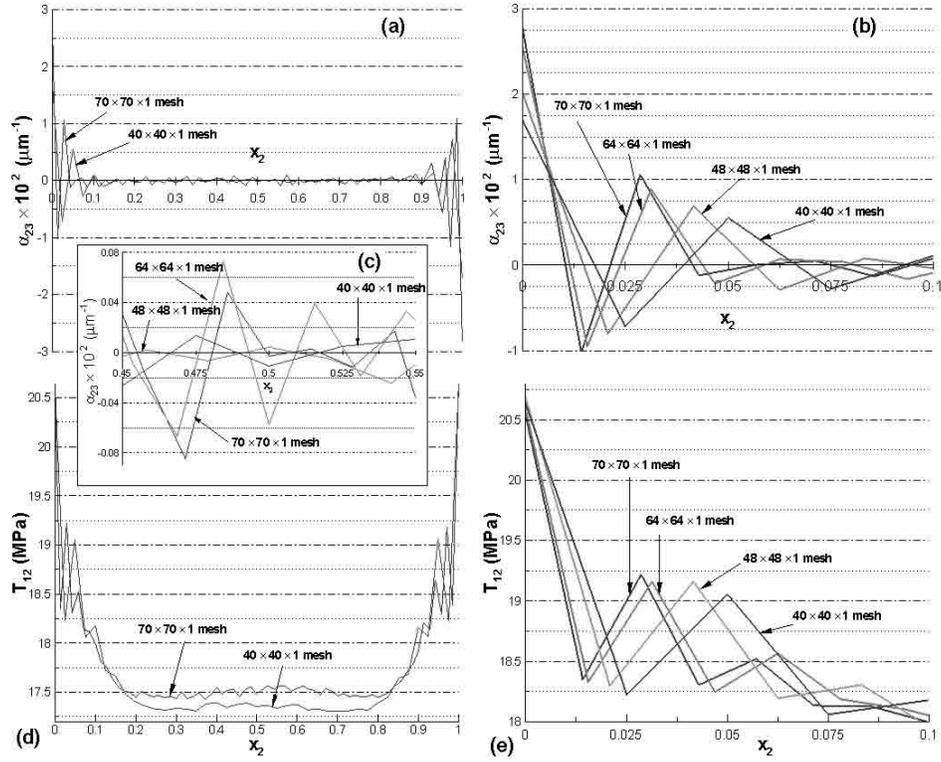

Fig. 10. Variation of fields along $x_2$ at $x_1 = 0.5\mu m$, $x_3 = 0.5\mu m$ of $(1\mu m)^3$ grain for different mesh sizes at $\Gamma = 0.2\%$. (a),(b),(c) variation of $\alpha_{23}$ (d),(e) variation of $T_{12}$

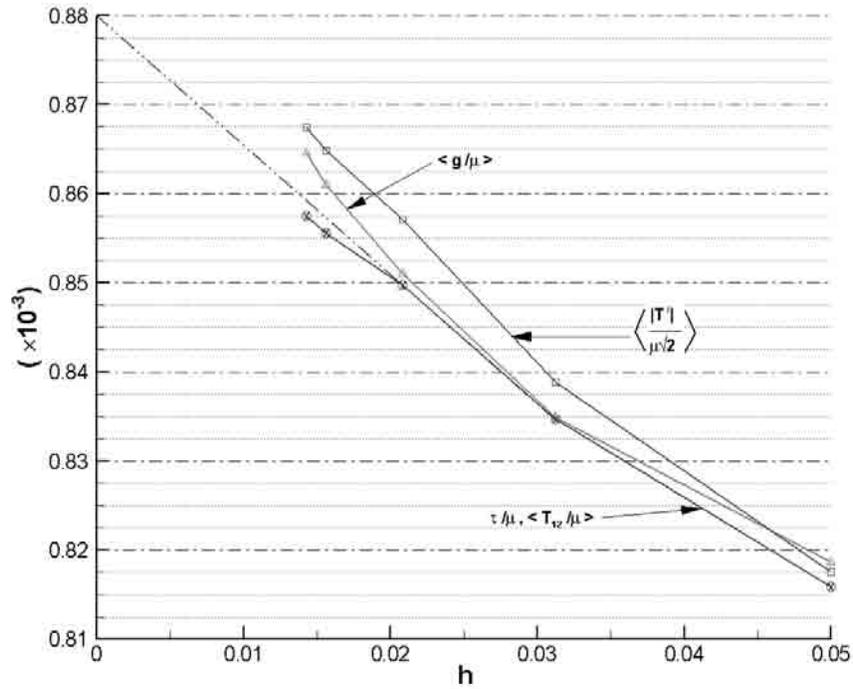

Fig. 11. Variation of top-layer average stress measures with element size, $h$, at $\Gamma = 0.2\%$. The notation $\langle * \rangle$ represents the average of $*$ over the top layer of elements.

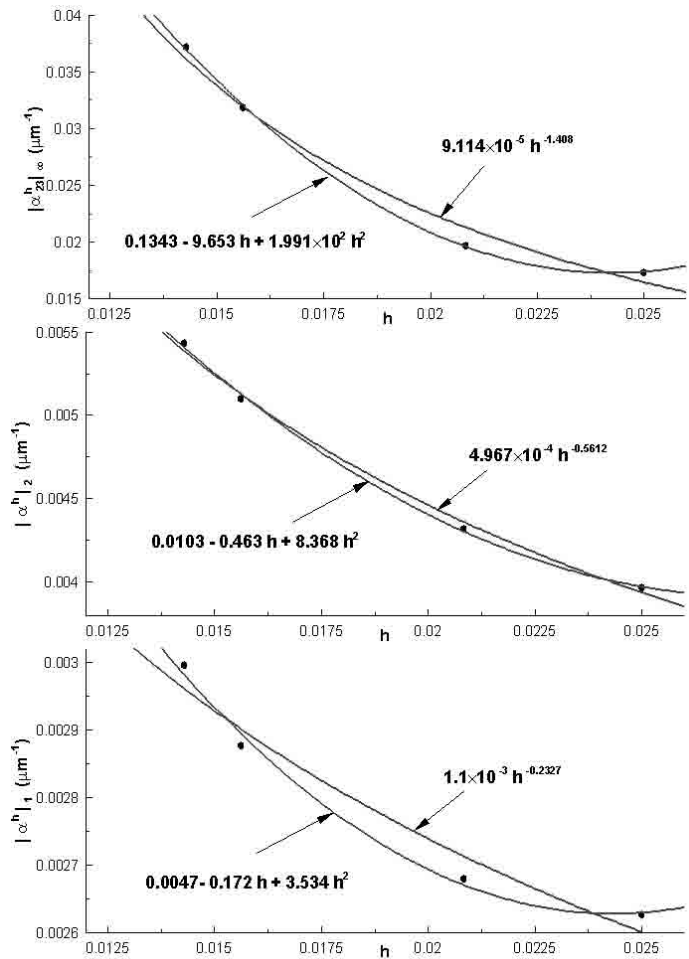

Fig. 12. Variation of $\alpha$ norms with element size $h$.

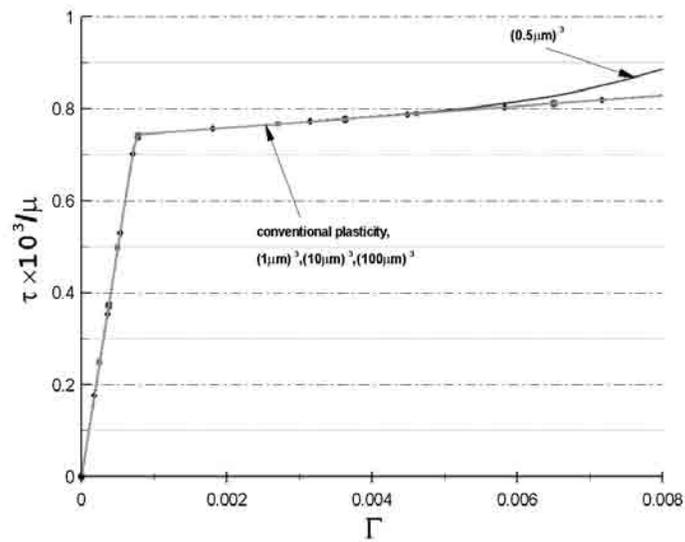

Fig. 13. Average $\tau - \Gamma$ response for different block sizes for control variables corresponding to homogeneous solution in conventional plasticity.

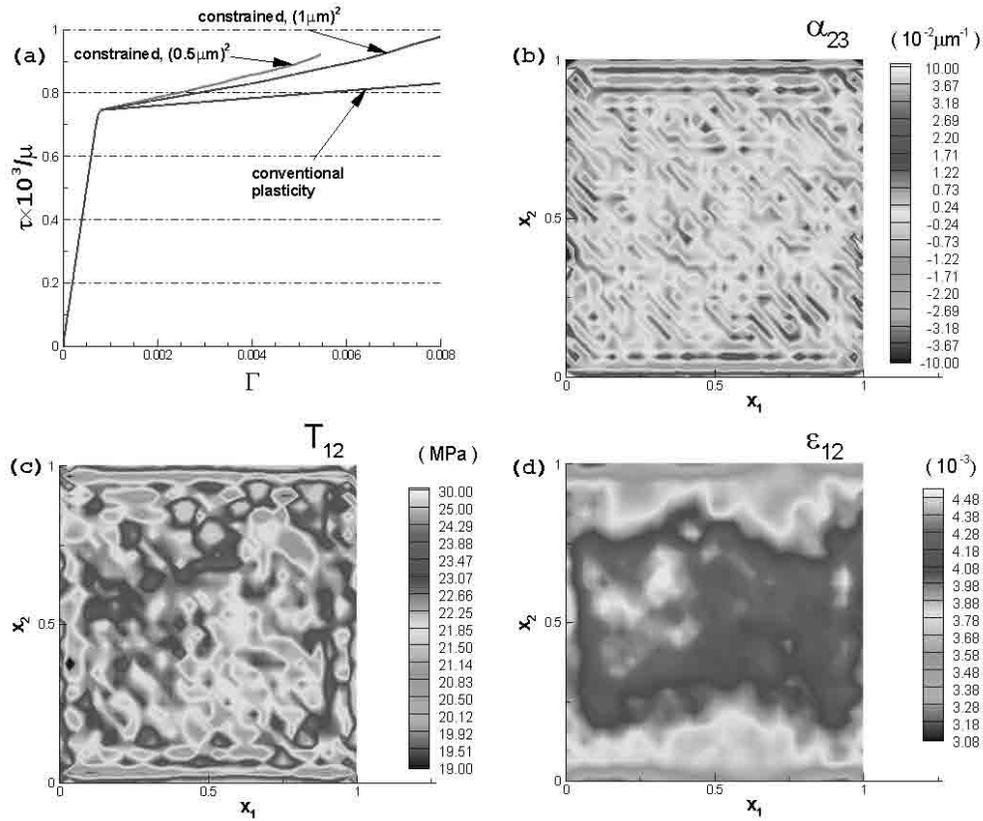

Fig. 14 (a) Size effect of infinitely long beam of different sizes. (b),(c),(d) $\alpha_{23}, T_{12}, \varepsilon_{12}$ profiles on the undeformed mesh at the end of loading for the larger film at $x_3 = 0.5\mu m$.

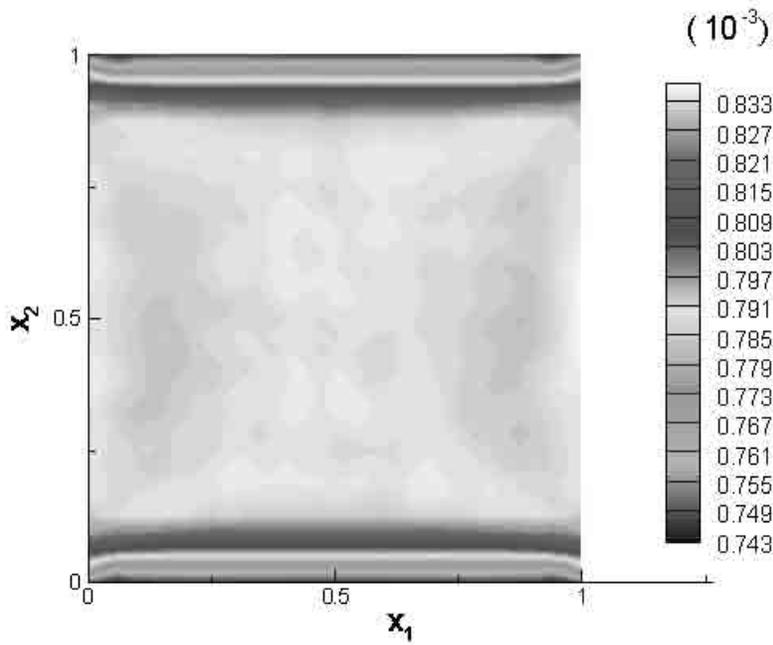

Fig. 15. $\varepsilon_{12}$ on the undeformed mesh at $\Gamma = 0.165\%$ for the bigger film at $x_3 = 0.5\mu m$.

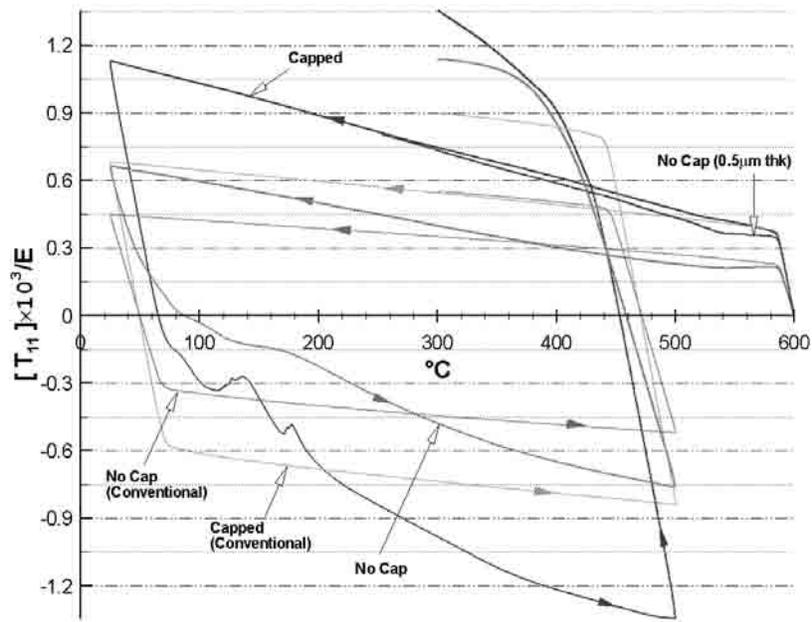

Fig. 16. Variation of $[T_{11}]$ with temperature for Capped, Uncapped films.

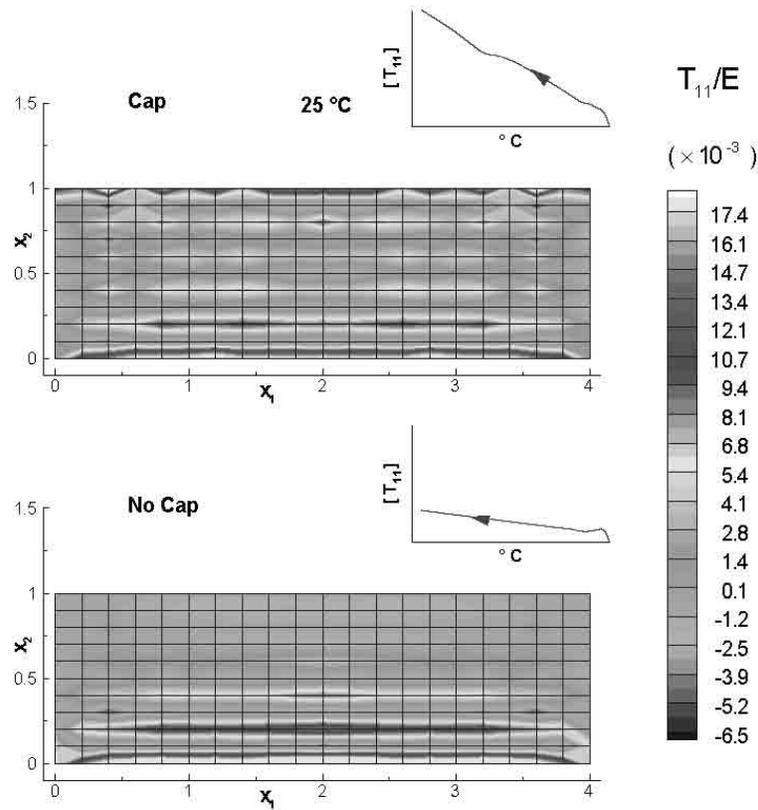

Fig. 17. $T_{11}$ profile for the $1\mu m$ thick capped and uncapped film at $25°C$. All plots are at $x_3 = 0.5\mu m$ on the undeformed mesh.

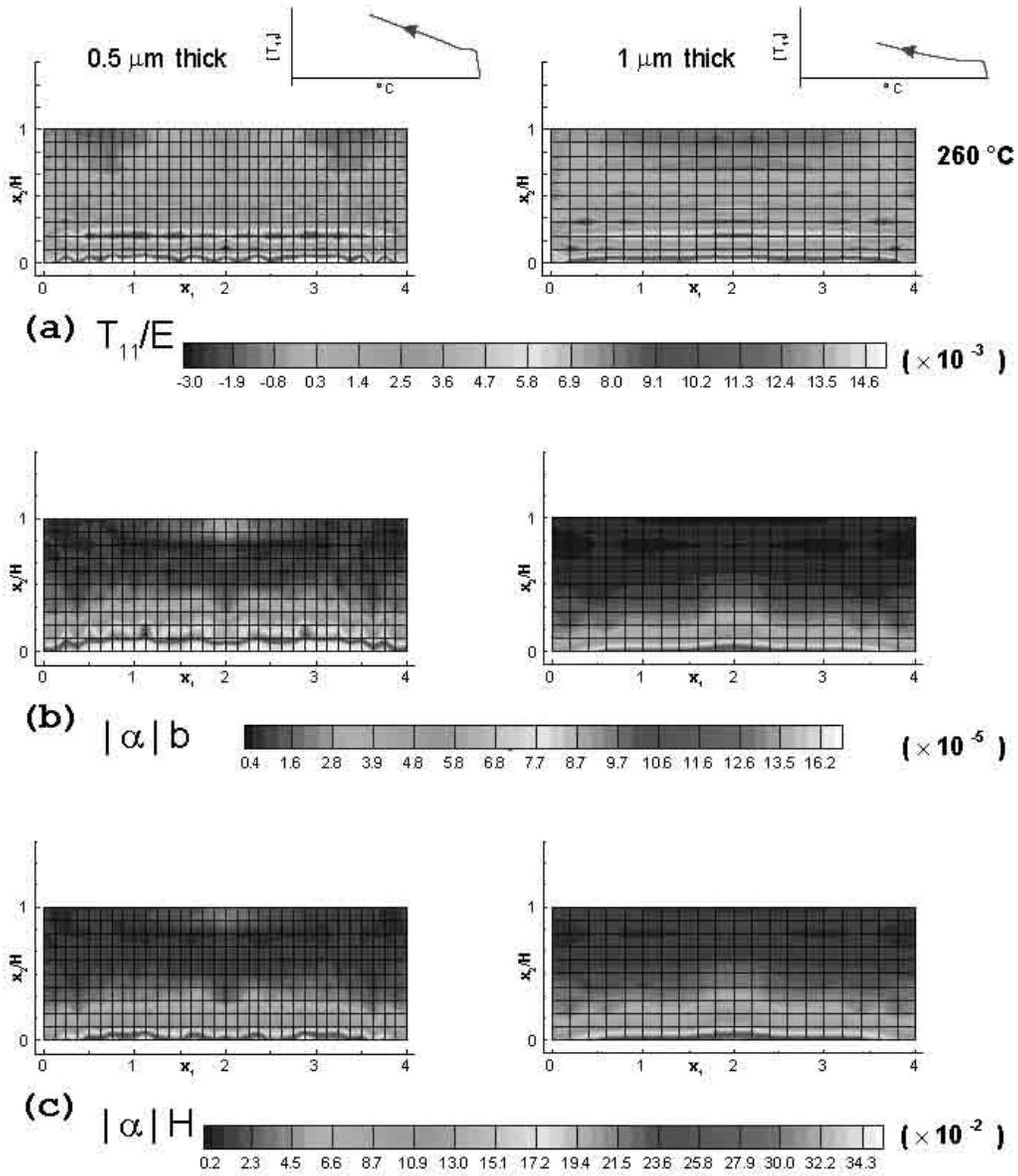

Fig. 18. Profiles for the $0.5\mu m$ and $1\mu m$ thick uncapped film at $260°C$. All plots are at $x_3 = 0.5\mu m$ on the undeformed mesh. (a) $T_{11}/E$ (b) $|\alpha|b$ (c) $|\alpha|H$, where $H$ is the film thickness.